\documentclass[aps,amsfonts,amsmath,prd,preprint,nofootinbib]{revtex4}
\usepackage[T1]{fontenc}
\usepackage[utf8]{inputenc}
\usepackage{array}
\usepackage{multirow}
\usepackage{amsmath}
\usepackage{amssymb}
\usepackage{graphicx}
\usepackage{esint}

\usepackage{stmaryrd}

\makeatletter

\providecommand{\tabularnewline}{\\}

\@ifundefined{textcolor}{}
{%
 \definecolor{BLACK}{gray}{0}
 \definecolor{WHITE}{gray}{1}
 \definecolor{RED}{rgb}{1,0,0}
 \definecolor{GREEN}{rgb}{0,1,0}
 \definecolor{BLUE}{rgb}{0,0,1}
 \definecolor{CYAN}{cmyk}{1,0,0,0}
 \definecolor{MAGENTA}{cmyk}{0,1,0,0}
 \definecolor{YELLOW}{cmyk}{0,0,1,0}
}

\@ifundefined{showcaptionsetup}{}{%
 \PassOptionsToPackage{caption=false}{subfig}}
\usepackage{subfig}
\makeatother

\begin{document}

\title{Primordial black hole formation by vacuum bubbles}

\author{Heling Deng and Alexander Vilenkin}

\affiliation{Institute of Cosmology, Tufts University, 574 Boston Ave, Medford,
MA, 02155 U.S.A.}
\begin{abstract}

Vacuum bubbles may nucleate during the inflationary epoch and expand, reaching relativistic speeds.   After inflation ends, the bubbles are quickly slowed down, transferring their momentum to a shock wave that propagates outwards in the radiation background.  The ultimate fate of the bubble depends on its size.  Bubbles smaller than certain critical size collapse to ordinary black holes, while in the supercritical case the bubble interior inflates, forming a baby universe, which is connected to the exterior region by a wormhole.  The wormhole then closes up, turning into two black holes at its two mouths.
We use numerical simulations to find the masses of black holes formed in this scenario, both in subcritical and supercritical regime.  The resulting mass spectrum is extremely broad, ranging over many orders of magnitude.  For some parameter values, these black holes can
serve as seeds for supermassive black holes and may account for LIGO observations.

\end{abstract}
\maketitle

\section{Introduction}

Primordial black holes (PBHs) are hypothetical black holes
formed in the early universe before any nonlinear large scale structure
and galaxies. The idea was conceived and developed decades ago \cite{Zeldovich,Hawking:1971ei,Carr:1974nx},
and since then PBHs have received considerable attention, despite
the fact that their existence is yet to be supported by observations.
Depending on the model, PBH masses can range from as low as the Planck
mass ($M_{\rm{Pl}}\sim10^{-5}$ g) to many orders of magnitude above
the solar mass ($M_{\odot}\sim10^{33}$ g). By contrast, black
holes formed by stellar collapse cannot have mass smaller than $M_{\odot}$.
Small PBHs $(M_{\rm{bh}}<10^{15}\text{ g})$ could be sources of
Hawking radiation, whereas PBHs with $M_{\rm{bh}}>10^{15}\text{ g}$
have been suggested as a candidate for (at least part of) the cold
dark matter and as possible seeds of supermassive black holes. Numerous
mechanisms of PBH formation have been proposed over the years. In
many scenarios (e.g., \cite{Carr:1993aq,GarciaBellido:1996qt,Yokoyama:1995ex,Frampton:2010sw,Clesse:2015wea}),
overdensity produced during inflation may overcome pressure and collapse
into a black hole after it reenters the horizon during the radiation-dominated
era. Other possibilities are related to first-order phase transitions
\cite{Kodama:1982sf,Jedamzik:1999am,Khlopov:1999ys}, the collapse
of cosmic string loops \cite{Hawking:1987bn,Polnarev:1988dh,Garriga:1992nm}
and domain walls \cite{Garriga:1992nm,Khlopov:2004sc,GVZ,DGV}, etc.

In this paper, we shall use numerical simulations to explore the possibility,
recently suggested in \cite{GVZ}, that PBH could be formed by nonperturbative
quantum effects in the early universe. Specifically, we will show
that spontaneous nucleation of vacuum bubbles during the inflationary
epoch can result in black holes with a wide mass spectrum at the present
time.

\begin{figure}
\includegraphics[scale=0.07]{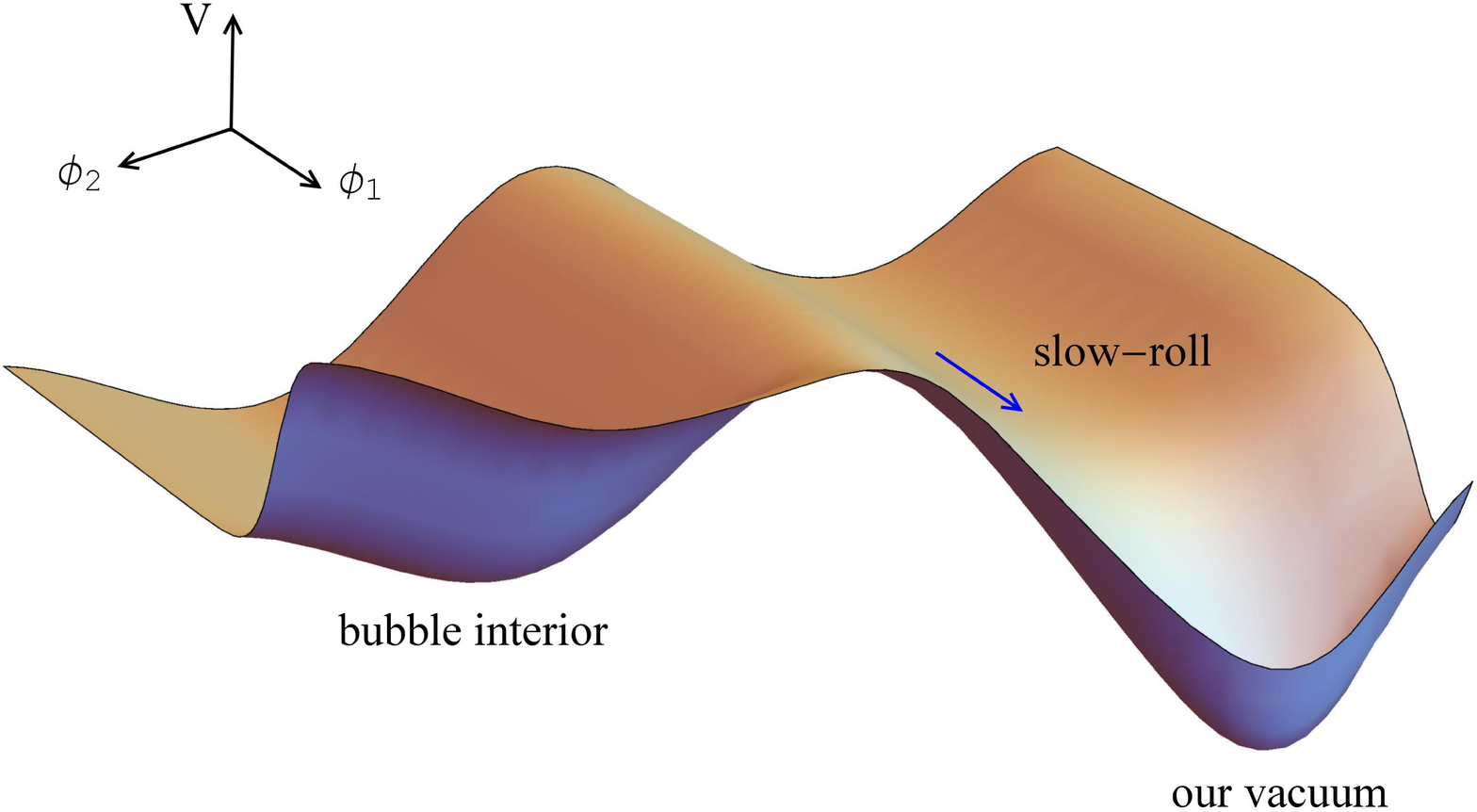}

\caption{\label{0001}A simple example of a two-field potential where the bubble nucleation scenario we discuss in this paper would be possible. As the inflaton field slowly rolls towards our vacuum, it can tunnel through a barrier to another vacuum, which will be the bubble interior.
}
\end{figure}

The physical mechanism responsible for this phenomenon is easy to
understand. The inflationary expansion of the universe is driven by
the high energy density $\rho_{i}$ of the false vacuum \footnote{We use the term \textquotedbl{}false vacuum\textquotedbl{} somewhat
loosely, including a slowly rolling inflaton in this category.}. Bubble nucleation may occur by quantum tunneling if the underlying
physics includes some vacuum states, other than our present vacuum,
with energy density $\rho_{\rm{b}}<\rho_{i}$ (we will be interested
in the case where $\rho_{\rm{b}}>0$). A two-dimensional energy landscape corresponding to this setup is illustrated in Fig. \ref{0001}.  The inflaton field slowly "rolls" along the gentle slope in the landscape towards a local energy density minimum representing our vacuum.  As it rolls, it can tunnel through a potential barrier to another vacuum of energy density higher than ours.  The tunneling occurs through bubble nucleation: a small spherical bubble of the new vacuum spontaneously forms in the inflating background \cite{Coleman:1980aw}. Once the bubble nucleates,
it expands with acceleration, acquiring a large Lorentz factor. This
growth of the bubble is caused by the large vacuum tension outside
(which is greater than the vacuum tension inside). At the end of inflation,
the false vacuum outside the bubble decays into hot radiation. The
bubble wall runs into the radiation and quickly loses much of its
energy, producing a shock wave that propagates outwards. A black hole
is then formed by one of the two different scenarios, depending on
the size of the bubble. (i) The bubble wall is pulled inwards by the
interior vacuum tension, the wall tension, as well as the radiation
pressure; so it shrinks and eventually collapses to a singularity.
Following \cite{GVZ}, we shall refer to such bubbles as subcritical.
(ii) If in the course of bubble expansion its size exceeds the interior
de Sitter horizon, the bubble begins to inflate. In the latter case,
the bubble continues to expand without bound and a wormhole is created
outside the bubble wall, connecting the inflating baby universe inside
and the FRW parent universe dominated by radiation. Such bubbles will
be called supercritical.

It was argued in Ref.~\cite{GVZ} that the bubble transfers most
of its kinetic energy to surrounding matter and comes to rest with
respect to the Hubble flow on a time scale much shorter than the Hubble
time. If the exterior region were filled with pressureless dust, an
empty layer would form around the bubble, so the bubble would be completely
isolated from matter. In this case the bubble evolution and the mass
of the resulting black hole can be found analytically. The case with
a radiation background is more involved, due to the pressure exerted
on the wall. Also, if a wormhole is formed, some radiation may follow
the bubble into the wormhole. Both of these effects may influence
the black hole mass.

Our goal in this paper is to numerically study black hole formation
by vacuum bubbles and to determine the resulting spectrum of black
hole masses. The paper is organized as follows. In Section II we discuss
the bubble dynamics in more detail and review some relevant earlier
work. Our simulation model is described in Section III, and simulation
results are presented in Section IV. We calculate the black hole mass
spectrum in Section V and discuss observational implications and constraints on the
model parameters in Section VI. Our conclusions are summarized and discussed in Section
VII.  We set $c=\hbar=1$ throughout the paper.

\section{Bubble dynamics}

We consider an idealized model where inflation ends instantaneously
at time $t=t_{i}$, so that false vacuum outside the bubble is instantly
turned into radiation of initial energy density $\rho_{i}$. We also
assume, as in Ref.~\cite{GVZ}, that particles are reflected from
the bubble wall, so radiation cannot penetrate the bubble and the
bubble interior always remains pure de Sitter.

As we mentioned in the Introduction, the bubble evolution
can lead to two possible outcomes, depending on the bubble
size. We first consider subcritical bubbles.

\subsection{Subcritical bubbles}

When the bubble transfers its momentum to radiation and
comes to rest with respect to the Hubble flow at $t\approx t_{i}$,
it continues to expand by inertia. But the forces due to the
tension of the vacuum inside the bubble and due to the tension of
the bubble wall are both directed inwards, so the bubble wall accelerates
inwards, away from the surrounding radiation, and we can expect that
in a few Hubble times the interaction of the bubble with radiation
becomes negligible.

\begin{figure}
\includegraphics[scale=0.11]{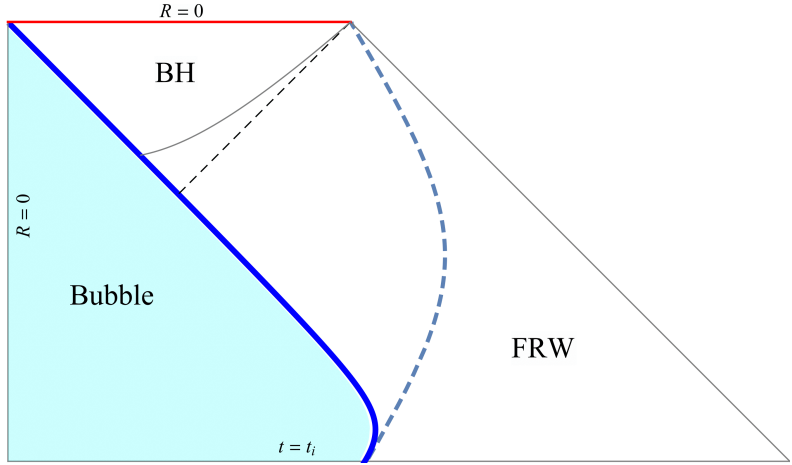}

\caption{\label{fig:sub}A conformal diagram showing the formation of
a black hole by a subcritical bubble in the background of a radiation
dominated spatially flat FRW universe. At the time $t_{i}$, when inflation ends, the bubble (the shaded region of the diagram) 
expands with a large Lorentz factor relative to the
Hubble flow. The bubble wall is represented by a thick blue solid curve.
The bubble expansion is slowed down by momentum transfer to the
ambient radiation, and eventually the bubble turns around 
and collapses into a Schwarzschild singularity
(red solid line). The thick dashed curve represents the shock front
propagating at the speed of sound, caused by the impact of the
fast-moving wall on the radiation. Region outside the shock front
is an unperturbed FRW universe. 
The spacelike curve below
the Schwarzschild singularity is the black hole apparent horizon,
which is used to represent the black hole boundary in our simulations.
It lies inside the event horizon (thin dashed straight line).}
\end{figure}

In the subcritical case, the bubble eventually stops expanding and
collapses to a Schwarzschild singularity; the corresponding conformal
diagram is shown in Fig. \ref{fig:sub}. We can estimate the mass
of the resulting black hole by assuming that the interaction with
radiation is negligible right after the initial momentum transfer at $t\approx t_{i}$. Then the bubble dynamics
is the same as it would be if the exterior region were asymptotically
Minkowski. In this case the bubble can be characterized by a conserved
mass parameter ${\cal M}_{\rm{b}}$ given by \cite{BKT} 
\begin{equation}
G{\cal M}_{\rm{b}}=\frac{1}{2}H_{\rm{b}}^{2}R_{\rm{w}}^{3}+2H_{\sigma}R_{\rm{w}}^{2}\sqrt{1+\dot{R}_{\rm{w}}^{2}-H_{\rm{b}}^{2}R_{\rm{w}}^{2}}-2H_{\sigma}^{2}R_{\rm{w}}^{3},\label{M_b}
\end{equation}
where $R_{\rm{w}}(\tau)$ is the bubble radius and the overdot stands
for a derivative with respect to the proper time $\tau$ on the bubble
wall. We have also defined\footnote{$H_{\rm{b}}$ is the rate of inflation in a vacuum of energy density $\rho_{\rm{b}}$, and $H_\sigma$ is the rate at which an isolated domain wall of tension $\sigma$ would inflate due to its self-gravity.} 
\begin{equation}
H_{\rm{b}}=\sqrt{\frac{8\pi G\rho_{\rm{b}}}{3}}
\end{equation}
and 
\begin{equation}
H_{\sigma}=2\pi G\sigma,
\end{equation}
where $\sigma$ is the bubble wall tension. The first term in Eq.
(\ref{M_b}) is the interior vacuum energy of the bubble, the second
term is the energy of the expanding wall, and the last term the gravitational
self-energy of the wall.

Since ${\cal M}_{\rm{b}}={\rm const}$, it can be evaluated at $t\approx t_{i}$,
when $\dot{R}_{\rm{w}}\approx H_{i}R_{i}$ with $H_{i}$ the Hubble
constant during inflation and $R_{i}$ the bubble radius at $t_{i}$.
The black hole mass is then simply $M_{\rm{bh}}={\cal M}_{\rm{b}}$.
Assuming that the bubble is much bigger than the horizon, $R_{i}\gg H_{i}^{-1}$,
this is given by 
\begin{equation}
GM_{\rm{bh}}\approx\left[\frac{1}{2}H_{\rm{b}}^{2}+2H_{\sigma}\left(\sqrt{H_{i}^{2}-H_{\rm{b}}^{2}}-H_{\sigma}\right)\right]R_{i}^{3}.\label{GM}
\end{equation}

The maximal radius of expansion of the bubble $R_{\rm{max}}$ can
be found by setting $\dot{R}_{\rm{w}}=0$ in Eq.~(\ref{M_b}) and
solving for $R_{\rm{w}}$. A solution exists only if $\mathcal{M}_{\rm{b}}$
is smaller than certain critical mass $M_{\rm{cr}}$. An exact expression
for $M_{\rm{cr}}$ was found in Ref.~\cite{Blau:1986cw}; it is
rather cumbersome and we will not reproduce it here. By order of magnitude,
$M_{\rm{cr}}$ can be estimated as \cite{GVZ} 
\begin{equation}
GM_{\rm{cr}}\sim\min\{H_{\rm{b}}^{-1},H_{\sigma}^{-1}\}.\label{Mcr}
\end{equation}

On dimensional grounds, $H_{\rm{b}}\sim\eta_{\rm{b}}^{2}/M_{\rm{Pl}}$
and $H_{\sigma}\sim\eta_{\sigma}^{3}/M_{\rm{Pl}}^{2}$, where $\eta_{\rm{b}}$
and $\eta_{\sigma}$ are the energy scales of the interior vacuum
and of the bubble wall, respectively. Then, with $\eta_{\rm{b}}\sim\eta_{\sigma}\ll M_{\rm{Pl}}$,
we have $H_{\rm{b}}\gg H_{\sigma}$ and $GM_{\rm{cr}}\sim H_{\rm{b}}^{-1}$.
In this case the maximum expansion radius is related to the black
hole mass as $GM_{\rm{bh}}\approx H_{\rm{b}}^{2}R_{\rm{max}}^{3}/2$.
As $M_{\rm{bh}}$ is increased, $R_{\rm{max}}$ grows and reaches
its largest value, $R_{\rm{max}}\sim H_{\rm{b}}^{-1}$ at the
critical mass. On the other hand, a situation where $\eta_{\rm{b}}\ll\eta_{\sigma}$
is also possible; then $H_{\sigma}\gtrsim H_{\rm{b}}$ and $GM_{\rm{cr}}\sim H_{\sigma}^{-1}$.
The choice of parameters in our simulations was dictated mostly by
the computing constraints.

We do not expect the estimates (\ref{M_b}) and (\ref{GM}) to be
very accurate. Radiation does work on the bubble while it expands
(which decreases the bubble mass) and while it contracts (which increases
the bubble mass). Since the radiation density is higher during the
expanding phase, one can expect the overall effect to be mass reduction
(except in cases where the expanding phase is very short). But since
the contact with radiation effectively ceases within a few Hubble
times, it was suggested in \cite{GVZ} that Eqs.~(\ref{M_b}) and
(\ref{GM}) should give the right order of magnitude. We shall see
that this is indeed the case.

\subsection{Supercritical bubbles}

\begin{figure}
\includegraphics[scale=0.11]{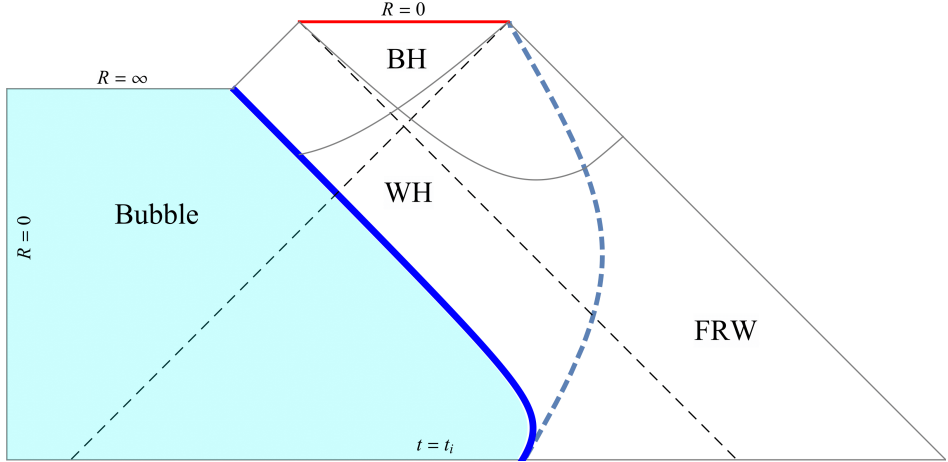}

\caption{\label{sup}A conformal diagram showing the formation of a black
hole by a supercritical bubble in a radiation dominated flat FRW universe. In this case, the bubble 
does not collapse into a singularity.
Instead, 
it grows exponentially in a baby
universe, which is connected by a wormhole to the parent FRW universe.
The thick dashed curve represents the shock front propagating at the
speed of sound, caused by the interaction of the fast-moving wall
and the radiation. Region outside the shock front is FRW dominated
by homogeneous radiation. The two intersecting spacelike curves below
the Schwarzschild singularity are the apparent horizons. The parts
above the intersection are black hole apparent horizons, representing
the boundary of two black holes. The right branch below the intersection
goes lightlike as it approaches the FRW lightlike infinity. This null
line is the Hubble radius (or cosmological apparent horizon) of the FRW universe. The two intersecting thin dashed straight lines below the
apparent horizons are the event horizons.
}
\end{figure}

For ${\cal M}_{\rm{b}}>M_{\rm{cr}}$, the bubble expands to a
radius greater than $H_{\rm{b}}^{-1}$, and the bubble interior
begins to inflate. At this point nothing can stop it from growing\footnote{For $H_{\sigma}>H_{\rm{b}}$, the bubble wall starts inflating,
due to its repulsive gravity, when its radius exceeds $H_{\sigma}^{-1}$.
Inflation in the bubble interior begins when the wall expands to $R_{\rm{w}}>H_{\rm{b}}^{-1}$.}. In this ``supercritical'' case, the bubble grows into a baby universe,
which is connected to the parent universe outside by a wormhole throat.
The wormhole closes up on a timescale $t\sim GM_{\rm{bh}}$, and
black holes of mass $M_{\rm{bh}}>M_{\rm{cr}}$ are formed at its
two mouths. From then on, the baby universe has no impact on further evolution of the exterior FRW region.  The corresponding
spacetime structure was discussed in Ref.~\cite{GVZ}; it is illustrated in the conformal diagram in Fig.
\ref{sup}.   As emphasized in \cite{GVZ}, wormhole formation in this spacetime does not violate any singularity theorems and does not require violation of the null energy condition.

An unusual feature of the diagram in Fig. \ref{sup} is the presence of a white hole region (marked WH), as in the Kruskal spacetime of an eternal black hole.  The boundaries of black and white holes are usually defined by their event horizons, and we adopted this convention in the figure.  A more physically motivated definition is to use apparent (or trapping) horizons (see e.g. Ref. \cite{VF} and references therein).  We followed this approach in our simulations; see Sec. III.D for more detail.

The shock wave produced by the bubble wall propagates outwards at
the speed of sound; its trajectory is shown by a dashed curve in the
diagram. The region outside the shock remains unperturbed and is described
by the FRW solution. On the other hand, the evolution inside the shock
radius can be rather complicated. In particular, some radiation flows out of the white hole region, resulting in a decrease of the black hole mass. This can be thought of
as an inverse process of radiation flowing into a black hole, which
would increase the mass.

A related problem has been studied in Ref.~\cite{DGV}, which investigated
the collapse of spherical domain walls formed during inflation. In
this case, inflation ends both inside and outside the wall. If the
wall expands to a radius greater than $H_{\sigma}^{-1}$, it starts
inflating and a radiation-filled baby universe is formed. It was pointed
out in \cite{DGV} that the Schwarzschild radius of the resulting
black hole cannot exceed the radius of the comoving FRW region affected by the
wall when it comes within the cosmological horizon. The affected region
is marked by the rarefaction wave that propagates away from the wall
at the speed of sound. The resulting bound on the black hole mass
is 
\begin{equation}
GM_{\rm{bh}}<2.8H_{i}R_{i}^{2}.\label{bound}
\end{equation}
The situation in our case is very similar, except the affected region
is now bounded by the shock front. Hence we expect the same bound
to apply. Numerical simulations in Ref.~\cite{DGV} showed that black
hole masses in supercritical regime are $GM_{\rm{bh}}\sim H_{i}R_{i}^{2}$,
so the bound (\ref{bound}) is nearly saturated. Here, we shall see
that the same conclusion applies to supercritical bubbles with a sufficiently large $R_{i}$.

\section{Simulation setup}

In this section we consider the equations of motion as well as the
initial and boundary conditions necessary for numerical simulations
of a vacuum bubble embedded in an otherwise homogeneous radiation-dominated
universe. We also indicate how we read the black hole mass from the
simulation results and discuss some simulation issues.

\subsection{Equations of motion and gauge conditions}

The spacetime is assumed to be spherically symmetric, and the metric
we use for the exterior region (outside of the bubble) is 
\begin{equation}
ds^{2}=-A^{2}dt^{2}+B^{2}dr^{2}+R^{2}d\Omega^{2},
\end{equation}
where $A,B$ and $R$ are functions of the coordinates $t$ and $r$.
The radiation fluid is generally described by its energy density $\rho$,
pressure $p=w\rho$ with $w=1/3$, and 4-velocity $u^{\mu}=(u^{0},u^{1},0,0)$
with $A^{2}(u^{0})^{2}-B^{2}(u^{1})^{2}=1$.

To fix the gauge, we choose the coordinates comoving with the fluid,
in which $u^{1}=0$ and $u^{0}=A^{-1}$. In this gauge, both the equations
of motion and the boundary conditions take a particularly simple form.
Since the fluid is confined to the bubble exterior, the normal component
of the fluid velocity vanishes at the bubble wall. The tangential
velocity vanishes by symmetry. It follows that the wall is comoving
with the fluid; hence it remains at a fixed value of the comoving
radius, $r=r_{\rm w}$. With a different choice of gauge we would
have to impose boundary conditions on a moving boundary, which is
considerably more complicated.

Following Ref.~\cite{Bloomfield} we introduce 
\begin{equation}
U\equiv\frac{\dot{R}}{A},\ \Gamma\equiv\frac{R^{\prime}}{B},\label{UGamma}
\end{equation}
where $\dot{}\equiv{\partial}/{\partial t}$ and $^{\prime}\equiv{\partial}/{\partial r}$.
Our goal is to solve Einstein's equations in order to find $A,U,\Gamma,B,R$
and $\rho$. By the following transformations 
\begin{equation}
\tilde{t}=H_{i}t,\ \tilde{B}=H_{i}B,\ \tilde{R}=H_{i}R,\ \tilde{\rho}=\frac{\rho}{M_{\rm{Pl}}^{2}H_{i}^{2}},\label{units}
\end{equation}
all variables become dimensionless. In this section we use these new
variables and drop the tilde. For instance, the time at the end of
inflation now becomes $t_{i}=(2H_{i})^{-1}=1/2$. To restore the physical
value of a certain quantity, one simply needs to multiply by an appropriate
conversion factor. For example, the conversion factor for mass is
$M_{\rm{Pl}}^{2}/H_{i}$.

Einstein's equations then take the form 
\begin{equation}
\frac{A^{\prime}}{A}=-\frac{w}{1+w}\frac{\rho^{\prime}}{\rho},\label{lapse}
\end{equation}
\begin{equation}
\dot{U}=-A\left(4\pi w\rho R+\frac{M}{R^{2}}\right)+\frac{A^{\prime}\Gamma}{B},\label{U}
\end{equation}
\begin{equation}
\dot{\Gamma}=\frac{A^{\prime}U}{B},\label{G}
\end{equation}
\begin{equation}
\dot{R}=AU.\label{dU}
\end{equation}
\begin{equation}
\dot{B}=\frac{AU^{\prime}}{\Gamma},\label{B1}
\end{equation}
\begin{equation}
\dot{\rho}=-(1+w)\rho A\left(\frac{U^{\prime}}{B\Gamma}+\frac{2U}{R}\right),\label{rho1}
\end{equation}
where 
\begin{equation}
M\equiv R(1-\Gamma^{2}+U^{2})/2
\end{equation}
is the Misner-Sharp mass parameter \cite{MS} that we shall use to
characterize the mass of the central object.

Eqs.~(\ref{B1}) and (\ref{rho1}) can be written in an equivalent
form: 
\begin{equation}
\dot{B}=\frac{AB}{U}\left(4\pi\rho R-\frac{M}{R^{2}}+\frac{\Gamma^{\prime}}{B}\right),\label{B2}
\end{equation}
\begin{equation}
\dot{\rho}=-(1+w)\frac{\rho A}{U}\left(4\pi\rho R-\frac{M}{R^{2}}+\frac{\Gamma^{\prime}}{B}+\frac{2U^{2}}{R}\right).\label{rho2}
\end{equation}
We used different equations in different situations in order to avoid
a vanishing denominator. For instance, in the subcritical case, the
bubble grows and then shrinks, so the value of $U$ at the wall goes
from positive to negative at the turning point. Hence we use Eqs.~(\ref{B1})
and (\ref{rho1}) to evolve $B$ and $\rho$ respectively, since $U=0$
must not appear in the denominator. Similarly, in the supercritical
case, $\Gamma$ crosses zero when the wormhole throat is formed, so
we use Eqs.~(\ref{B2}) and (\ref{rho2}).

The gauge condition $u^{1}=0$ leaves the freedom of time transformations,
$t\to{\bar{t}}(t)$. We can fix the gauge completely by specifying
$A(r,t)$ on any timelike curve. Before the bubble is removed for
reasons discussed below (Subsection IV.B), a convenient choice is to set $A(r_{\rm w},t)=1$
at the bubble wall. Then our time coordinate $t$ coincides with the
proper time $\tau$ at the wall.

\subsection{Boundary conditions}

We first comment on the number of boundary conditions required for
our problem. The following argument was suggested to us by Andrei
Gruzinov.

Introducing two new variables, $C=A^{3}/B$ and $F=\Gamma^{2}-U^{2}$,
instead of $B$ and $\Gamma$, the equations of motion (\ref{U})-(\ref{rho1})
can be represented as 
\begin{equation}
{\dot{R}}=...~,~~~{\dot{C}}=...~,~~~{\rm{\ensuremath{\dot{F}}}}=...~,\label{RCF}
\end{equation}
\begin{equation}
{\dot{U}}=-\frac{w}{1+w}\frac{A\Gamma}{B\rho}\rho'+...~,\label{Urho}
\end{equation}
\begin{equation}
{\dot{\rho}}=-(1+w)\frac{A\rho}{B\Gamma}U'+...~,\label{rhoU}
\end{equation}
where \textquotedbl{}...\textquotedbl{} means terms without derivatives
and we have used Eq.~(\ref{lapse}) to express $A'$ in terms of
$\rho'$ in (\ref{Urho}). Now, Eqs.~(\ref{RCF}) do not require
boundary conditions (only initial conditions) and Eqs.~(\ref{Urho})
and (\ref{rhoU}) represent a wave with a (high-frequency) sound speed
$c_{s}^{2}=wA^{2}/B^{2}$. We thus have one propagating degree of
freedom, which requires one left and one right boundary condition.

Variables outside of the shock front should be described by the unperturbed
FRW solution, so the outer boundary condition is easy to impose. Before
the bubble is removed, the wall serves as the inner boundary. The
bubble interior is described by de Sitter space with energy density
$\rho_{\rm{b}}$. The boundary conditions at the wall can be obtained
using Israel's junction conditions. This is done in Appendix A.\footnote{The equation of motion of the wall is also derived in Appendix A, but we did not use it in our simulations.}  
As we discussed, we only need one boundary
condition at $r=r_{\rm w}$; we use the condition \eqref{Aprime-1}:
\begin{equation}
A^{\prime}=-AB\left(\frac{w\rho+\rho_{\rm{b}}}{\sigma}+\frac{2\Gamma}{R}+6\pi\sigma\right).\label{Aprime}
\end{equation}
We also set $A=1$ on the inner boundary at all times, even after
the bubble wall is removed.

\subsection{Initial conditions}

We assume an idealized initial state where inflation ends instantaneously
at $t=t_{i}$ and false vacuum energy immediately turns into radiation
of uniform energy density 
\begin{equation}
\rho_{i}=\frac{3}{32\pi t_{i}^{2}}=\frac{3}{8\pi}.\label{rhoi}
\end{equation}
Since the lapse function $A=1$ at $r=r_{\rm w}$, it follows from
Eq.~(\ref{lapse}) that $A(r,t_{i})=1$ in the entire region outside
of the bubble. One might expect that at $t=t_{i}$ this whole region
is described by the FRW solution. Then, with a suitable normalization
of $r$, we would have $B(r,t_{i})=1$, $R(r,t_{i})=r$, $U(r,t_{i})=r/2t_{i}$,
and $\Gamma(r,t_{i})=1$. However, these values cannot be imposed
as initial conditions in the entire region $r>r_{\rm w}$.

At the initial moment, the bubble wall moving with a large Lorentz
factor comes into contact with the radiation fluid that
surrounds it. On the other hand, we are working in a gauge where
the radiation is comoving with the wall at $r=r_{\rm w}$.
Clearly, this condition cannot be satisfied if the FRW solution outside
the bubble is unperturbed. We deal with this problem by modifying
the FRW solution in a thin layer around the bubble.

The Misner-Sharp mass of the bubble at $t_i$ is $M(r_{\rm w},t_i)=H_i^2R_i^3/2=r^3_{\rm w}/2$, where we have assumed that  $R_i=r_{\rm w}$. On the other hand, $M(r_{\rm w},t_i)$ is also given by Eq. \eqref{M_b} with $R_{\rm w}=R_i$. Then by the definition of $U$ and $M$, the initial value of $\Gamma$ at the wall is given by 
\begin{equation}
\Gamma(r_{\rm w},t_{i})=\left(\frac{1-H_{\rm{b}}^{2}}{4H_{\sigma}}-H_{\sigma}\right)r_{\rm w}.\label{Gammain}
\end{equation}
To smooth out the discontinuity between this and the FRW value of
$\Gamma=1$, we use the following function for the initial profile
of $\Gamma$, 
\begin{equation}
\Gamma(r,t_{i})=\left[1-\Gamma(r_{\rm w},t_{i})\right]\tanh\left(\frac{r-r_{\rm w}}{\delta}\right)+\Gamma(r_{\rm w},t_{i}), \label{Gammain2}
\end{equation}
where $\delta$ characterizes the thickness of the layer that connects
the wall and the FRW universe.

\begin{figure} \includegraphics[scale=0.096]{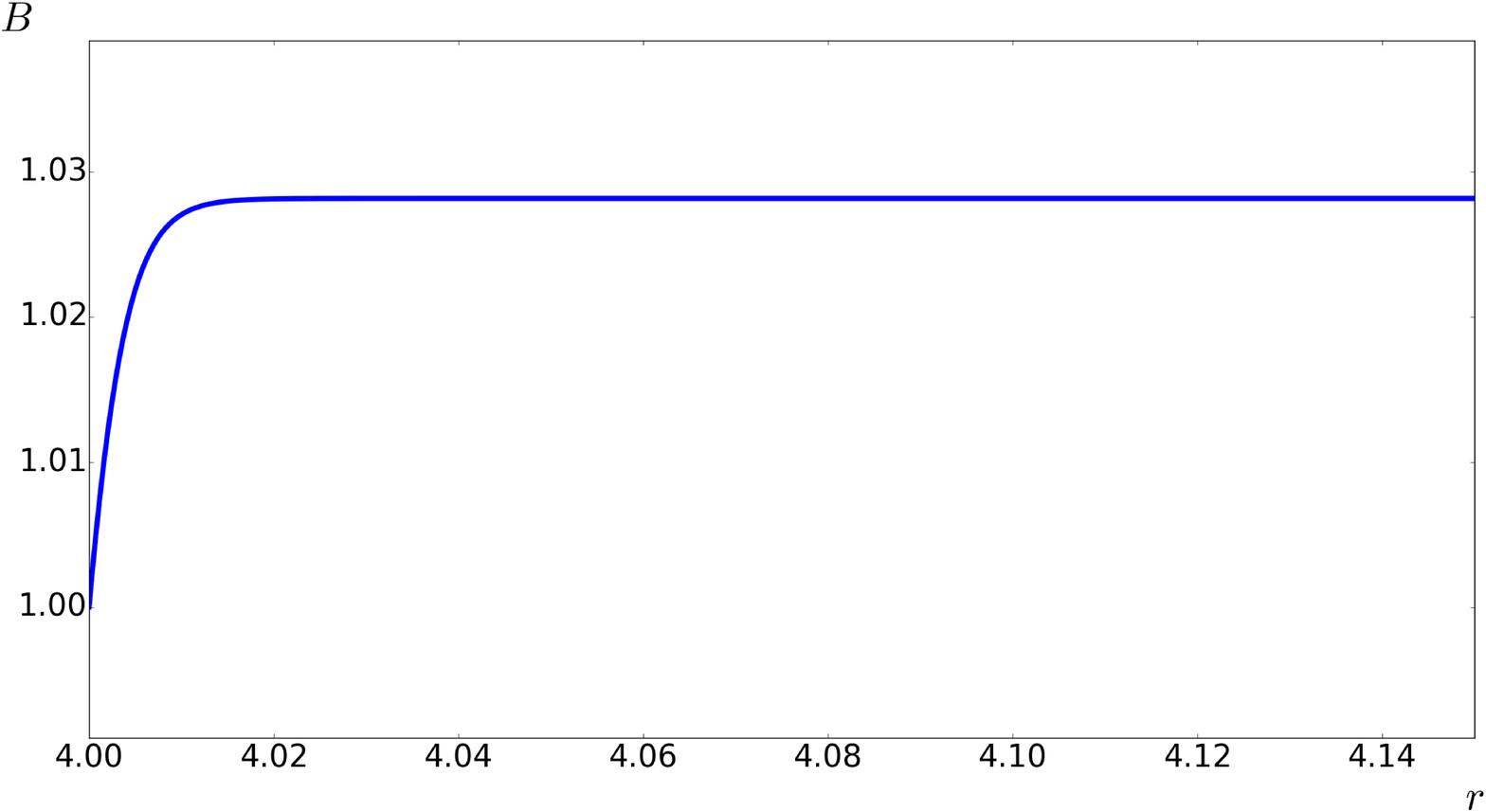}\includegraphics[scale=0.096]{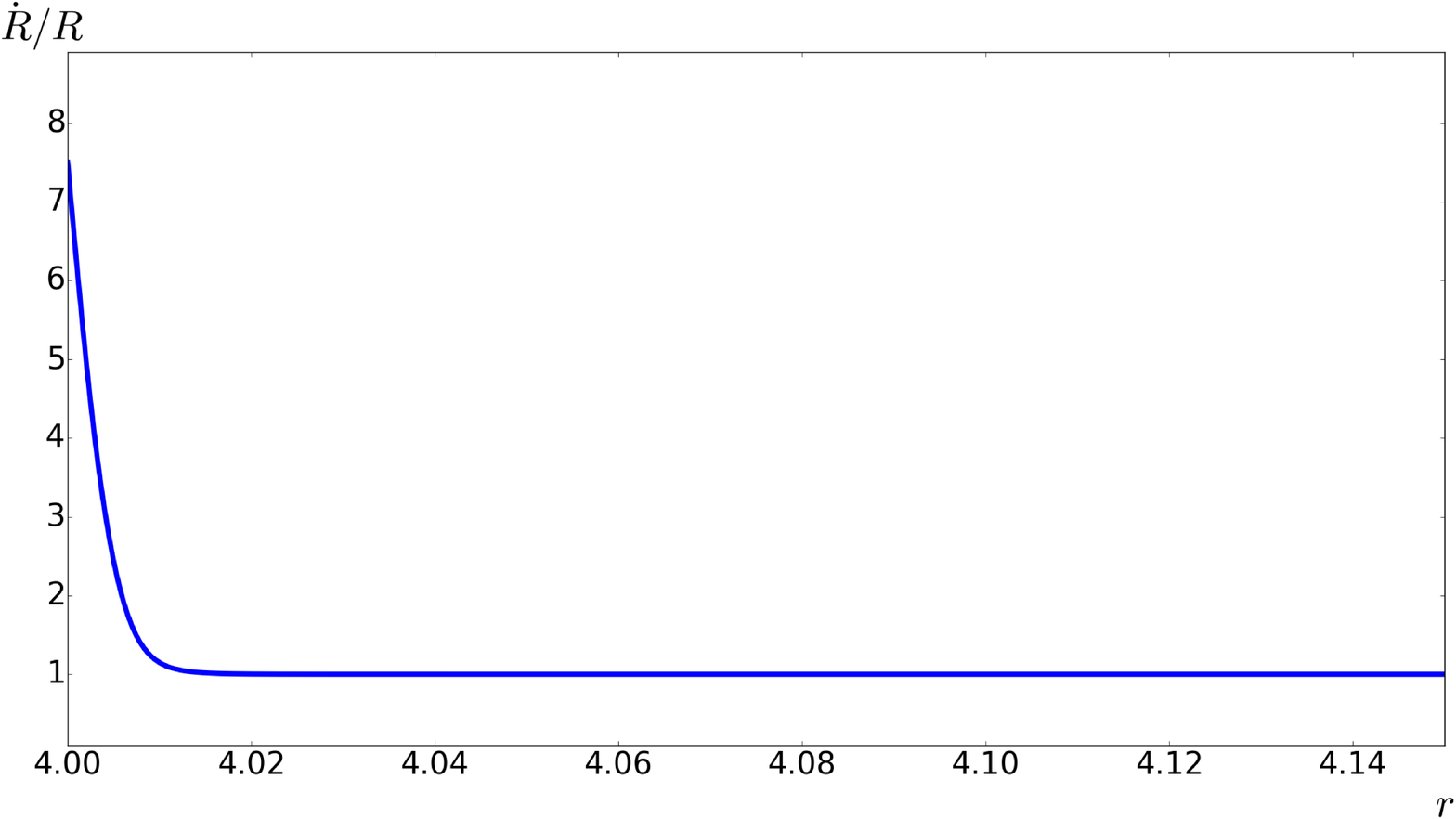}

\caption{\label{fig:The-initial-profiles}The initial profiles of $B$ and
$\dot{R}/R$ for a bubble with $R_{i}=4.$}
\end{figure}

To fix the remaining initial conditions, we assume the spatial metric $ds^{2}=B^{2}(dr^{2}+r^{2}d\Omega^{2})$.
Then $R(r,t_{i})=B(r,t_{i})r$ and it follows from the definition
of $\Gamma$ that 
\begin{equation}
\frac{R^{\prime}r}{R}=\Gamma.\label{BGamma}
\end{equation}
$R(r,t_{i})$ can now be found by numerically integrating Eq. (\ref{BGamma}). To illustrate the deviation of our initial state from FRW, we plot
the functions $B(r,t_{i})$ and $\dot{R}(r,t_{i})/R(r,t_{i})$ in
Fig. \ref{fig:The-initial-profiles}. (Both of these functions are
equal to 1 in an FRW universe.)

The Misner-Sharp mass within a radius\textbf{ $r$} at $t_{i}$
can be found from the relation \cite{hayward} $M^{\prime}=4\pi\rho R^{2}R^{\prime}$,
which gives $M(r,t_{i})=R^{3}(r,t_{i})/2$. Then by the definition of $M$, the initial profile of $U$ is
\begin{equation}
U(r,t_{i})=\sqrt{R^{2}+\Gamma^{2}-1}.
\end{equation}

It can be shown \cite{GVZ} that in our gauge the initial value of $U$ at the wall is related to the initial Lorentz factor of the wall $\gamma$ (relative
to an FRW observer) by
\begin{equation}
U(r_{\rm w},t_{i})=R_{i}\gamma+\sqrt{\gamma^{2}-1}.\label{gamma}
\end{equation}
In our simulations, we used $H_{\rm{b}}$ and $\gamma$ as free
parameters. The value of $H_{\sigma}$ can be determined from Eqs.
\eqref{M_b}, \eqref{UGamma} and \eqref{gamma}.

A profile of $U(r,t_{i})$ with a finite $\delta$ means that the
fluid already acquired some kinetic energy at $t_{i}$ in a layer
of width $\delta$. It is shown in Ref.~\cite{GVZ} that the wall
loses most of its kinetic energy and comes to rest with respect to
the radiation fluid within a time $\Delta t\sim H_{\sigma}^{2}/H_{i}^{3}$
(a more accurate calculation gives $\Delta t\sim100H_{\sigma}^{2}/H_{i}^{3}$).
This suggests that $\delta$ should be chosen so that $\delta\lesssim\Delta t$.
Indeed, we have verified that reducing $\delta$ below this value
does not have a significant impact on the black hole mass.

\subsection{Expansions and horizons}

We determine the formation of a black hole by checking if an apparent
horizon bounding a trapped region is formed. Let $\Theta_{\rm{out}}$
and $\Theta_{\rm{in}}$ be the expansions of outgoing and ingoing
radial null geodesics respectively; then a surface is trapped if $\Theta_{\rm{out}},\Theta_{\rm{in}}<0$,
and anti-trapped if $\Theta_{\rm{out}},\Theta_{\rm{in}}>0$. In
our coordinate system \cite{numerical R}, 
\begin{equation}
\Theta_{\rm{out}}\propto\frac{U+\Gamma}{R},\ \Theta_{\rm{in}}\propto\frac{U-\Gamma}{R}.\label{theta}
\end{equation}

The apparent horizon of a black hole is foliated by marginal spheres
with $\Theta_{\rm{out}}=0$ and $\Theta_{\rm{in}}<0.$ We also
define the white hole (apparent) horizon and cosmological (apparent)
horizon to be hypersurfaces foliated by marginal spheres with $\Theta_{\rm{in}}=0$
and $\Theta_{\rm{out}}>0.$ A spherical surface is anti-trapped
within the white hole horizon, and is normal (with $\Theta_{\rm{in}}>0$
and $\Theta_{\rm{out}}>0$) if it lies between the white (or black
hole) and the cosmological horizon. The cosmological horizon and the
Hubble radius coincide in a flat FRW universe.

\subsection{Simulation issues}

We use finite-difference method and Runge-Kutta integration to solve
the PDEs.

Special attention is needed on the shock wave. The shock formed by
the interaction of the fast-moving wall and the radiation may lead
to numerical instability. A standard and convenient trick to handle
this is to introduce artificial viscosity \cite{VonNeumann} in order
to smooth out the discontinuity. Following \cite{LRT}, we add an
extra term to $w$, 
\begin{equation}
w\to\frac{1}{3}+\beta\Delta r^{2}\left(\frac{U^{\prime}}{R^{\prime}}+\frac{2U}{R}-\left|\frac{U^{\prime}}{R^{\prime}}+\frac{2U}{R}\right|\right)\left(\frac{U^{\prime}}{R^{\prime}}-\frac{U}{R}\right),\label{vis}
\end{equation}
where $\beta$ is an adjustable coefficient that controls the viscosity
strength, and $\Delta r$ is the grid size.
When needed, we replace $U^{\prime}/R^{\prime}$ by 
\begin{equation}
\frac{U^{\prime}}{R^{\prime}}\to U^{-1}\left(4\pi\rho R-\frac{M}{R^{2}}+\frac{\Gamma^{\prime}}{B}\right)
\end{equation}
to avoid $R^{\prime}=0$ in the denominator in Eq. (\ref{vis}).

Additionally, in order to improve the efficiency of the code, we use
an adaptive non-uniform mesh. At the beginning of the simulation,
a sufficiently high resolution is used until a shock wave is formed.
Then we reduce the mesh density in regions far away from the shock.
We keep track of the shock and make sure there is a sufficient number
of grid points there.

Another issue is related to the early evolution of the bubble wall.
A very large initial Lorentz factor $\gamma$ tends to break down
the code before any desirable results are obtained. Throughout the
simulations we used $\gamma\lesssim10$. This leads to a constraint
on $H_{\sigma}$. The initial Lorentz factor can be estimated as \cite{GVZ}
$\gamma\sim H_{\sigma}^{-1}$. Using this and Eqs.~(\ref{GM}) and
(\ref{Mcr}), it can be shown that the radii of subcritical bubbles
must satisfy $R_{i}\lesssim\gamma$. On the other hand, 
we are mostly interested in bubbles of initial radius
greater than the horizon, $R_{i}\gtrsim1$.
This gives a rather limited range of values for $R_{i}$. For supercritical
bubbles, we are free to use a large value for $R_{i}$, but are restricted
by the simulation runtime.

\section{Simulation results}

\subsection{Shock propagation}

\begin{figure}
\subfloat[]{\includegraphics[scale=0.095]{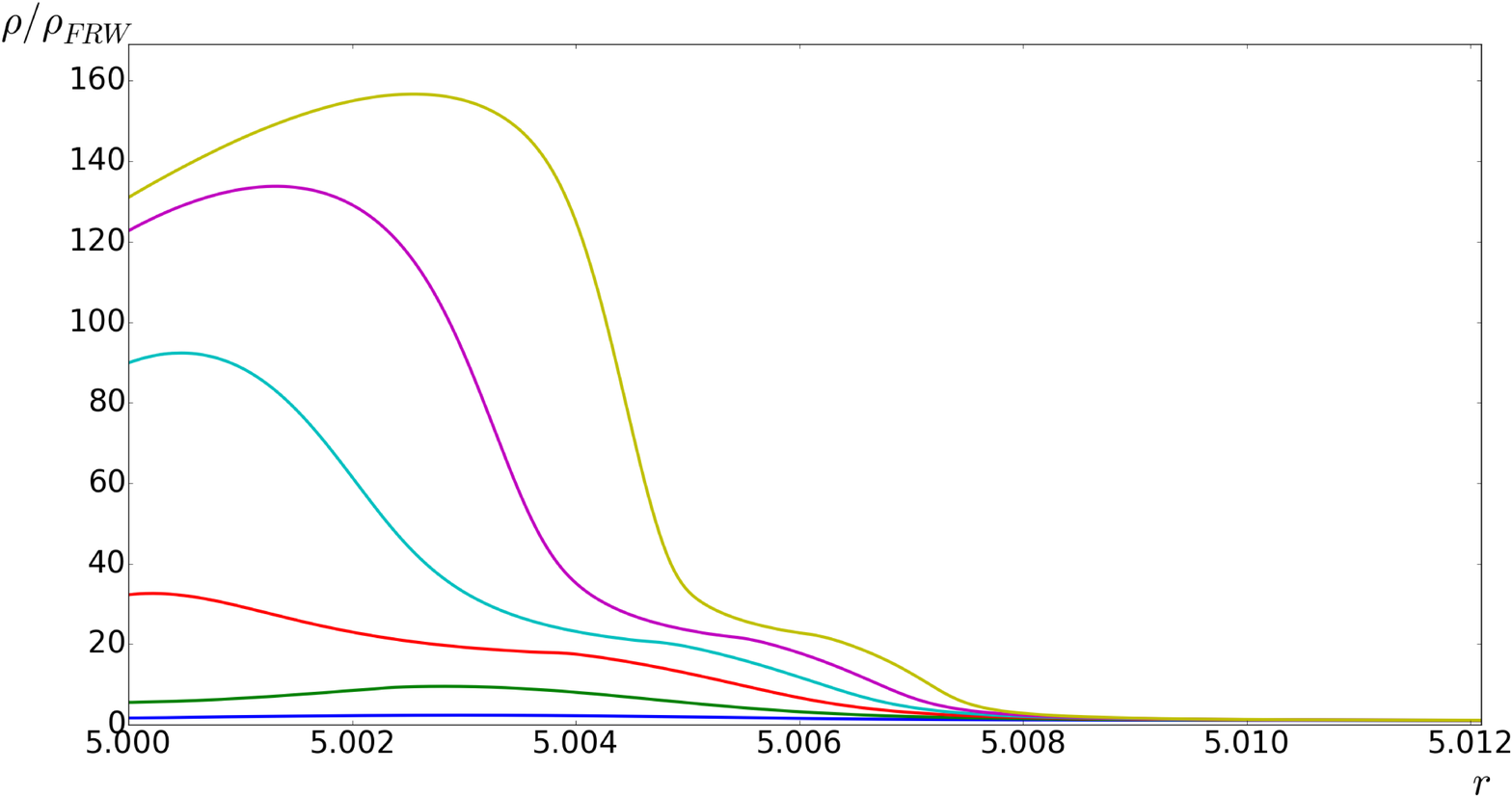}

}\subfloat[]{\includegraphics[scale=0.095]{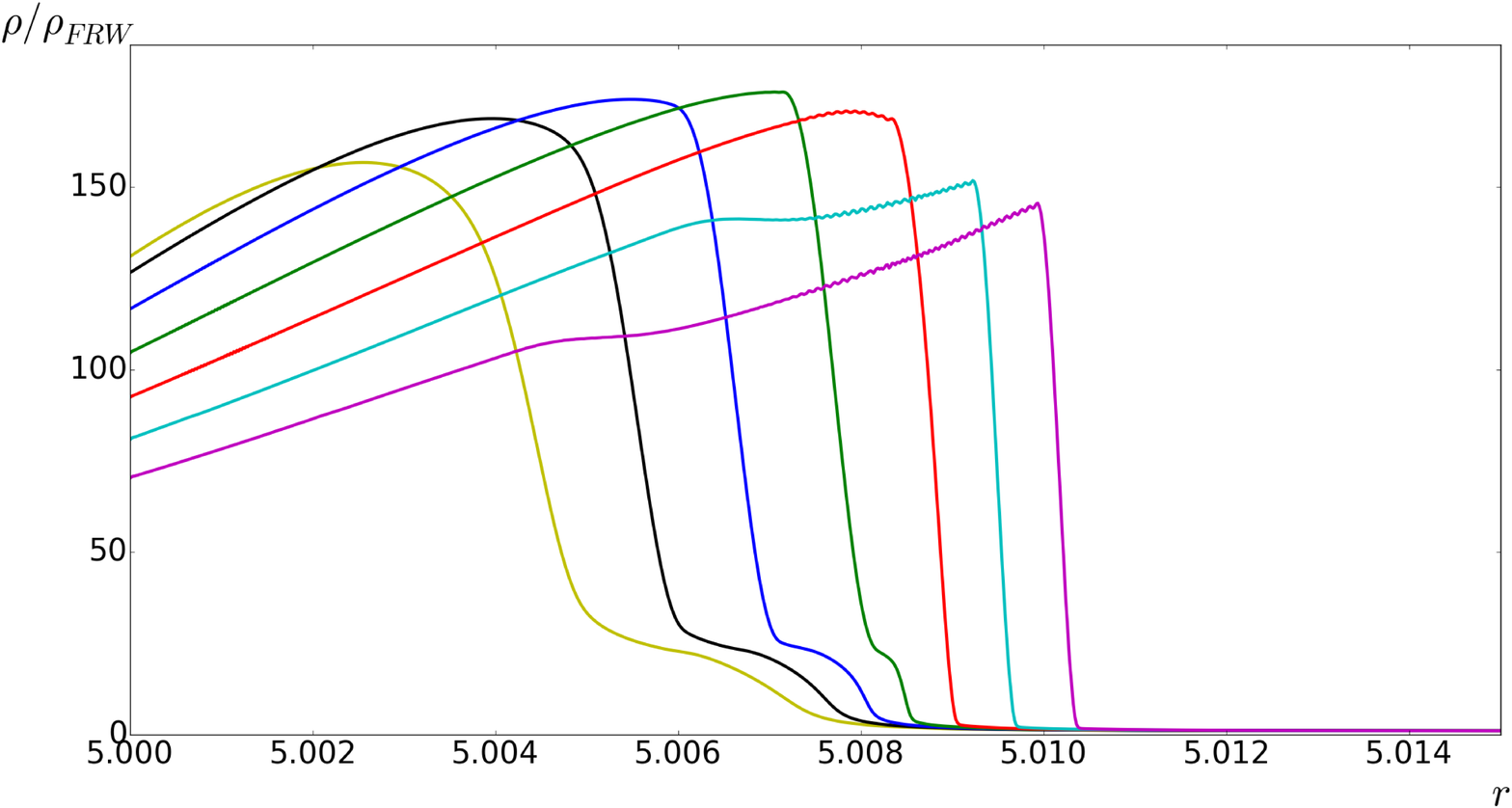}

}

\subfloat[]{\includegraphics[scale=0.095]{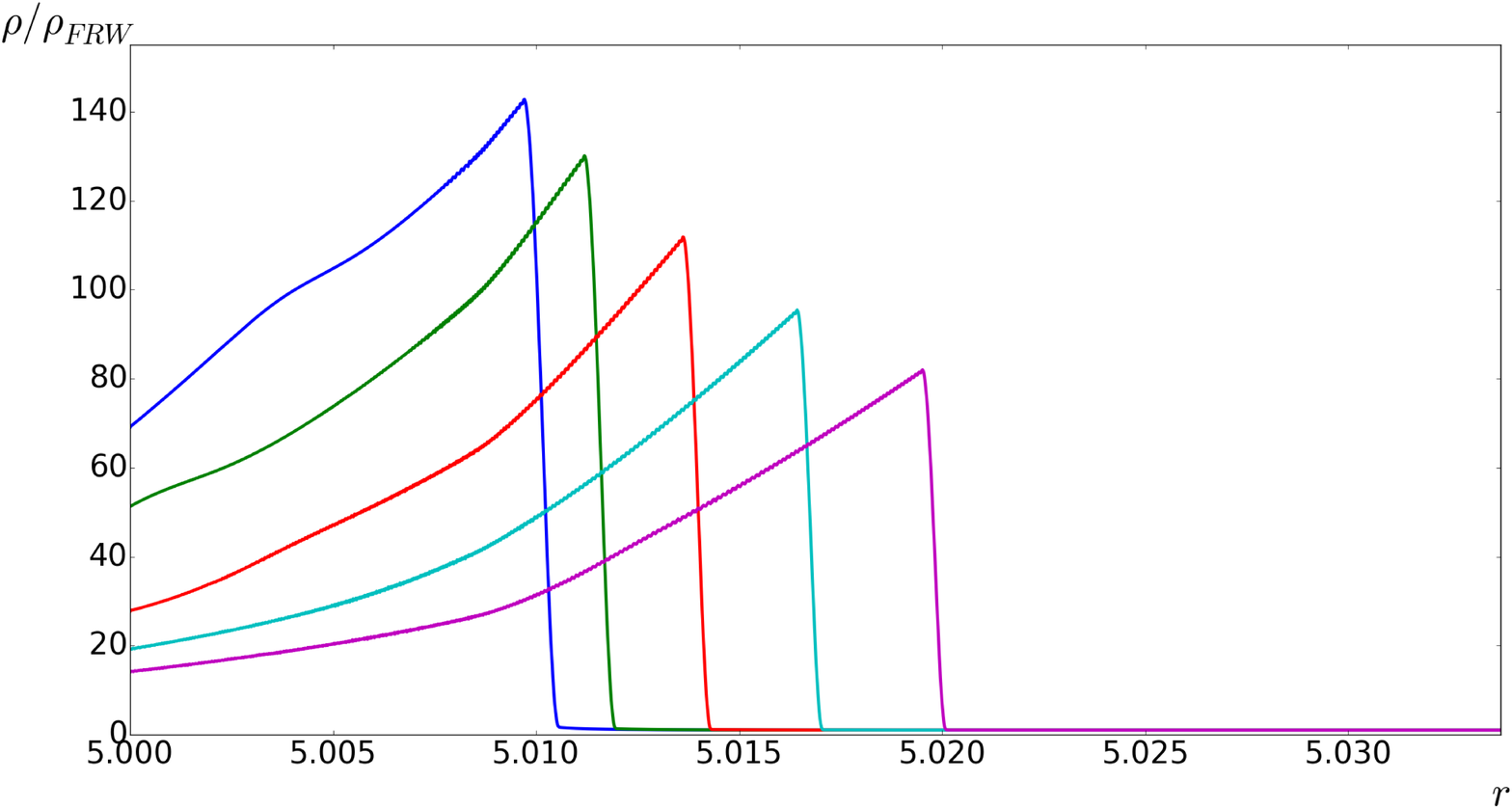}

}\subfloat[]{\includegraphics[scale=0.095]{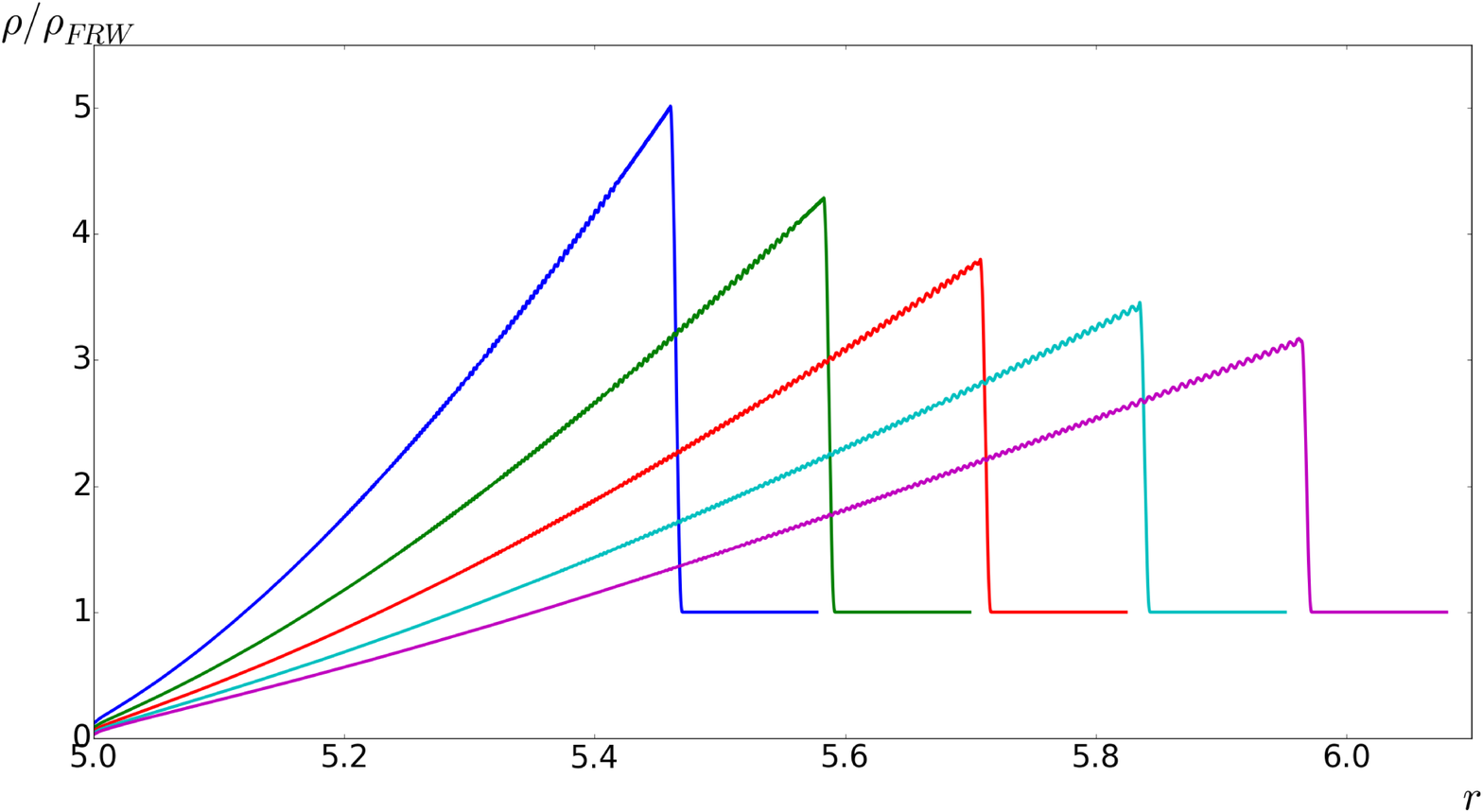}

}

\caption{\label{fig:The-radiation-energy}The radiation energy density $\rho$
as a function of the comoving radius $r$ at different moments of
time outside of a subcritical bubble with $H_{\rm{b}}=0.05H_{i}$,
$H_{\sigma}\approx0.03H_{i},$ and $R_{i}=5H_{i}^{-1}$. For all moments,
$\rho$ has been rescaled so that the FRW density is 1. (a), (b) and
(c) are taken at FRW times $t$ when $t-t_{i}\ll t_{i}$, while (d)
is at $t$ when $t-t_{i}\sim t_{i}$. (a) An overdense layer is formed
next to the wall as it hits the fluid. (b) A shock wave forms and
propagates outwards, while the density at the wall begins to decrease.
(c) The shock continues to propagate with the density contrast across
the shock rapidly decreasing. (d) $\rho$ right outside the wall becomes
much smaller that the FRW density, as if the bubble is surrounded
by an empty layer, much like in the case of a dust background. }
\end{figure}

\begin{figure}
\includegraphics[scale=0.17]{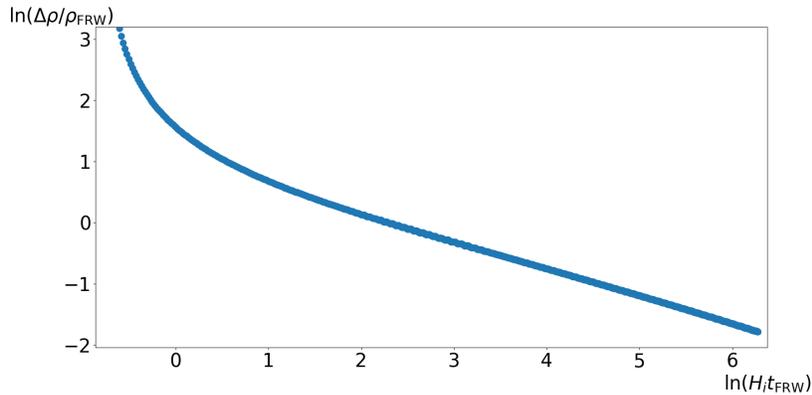}

\caption{\label{fig:The-density-contrast}The density contrast across the shock
$\delta_{s}$ as a function of time for the subcritical bubble
in Fig \ref{fig:The-radiation-energy}. As the shock propagates outwards,
it approaches the speed of sound and satisfies $\delta_{s}(t)\propto t^{\epsilon}$,
where $\epsilon\approx-1/2$. In the example shown here $\epsilon\approx-0.46$.}
\end{figure}

The formation and propagation of the shock wave produced by the bubble
are illustrated in Fig. \ref{fig:The-radiation-energy}. The figure
shows the radiation density profile at successive moments of time.
As the bubble wall hits the ambient radiation, a thin overdense layer
is formed right outside the wall. In this example, the density in
the layer exceeds that in the FRW region by more than 100 times. The
overdense layer spreads and develops a sharp shock front, which then
propagates outwards. Meanwhile, the radiation density next to the
wall rapidly drops, and in less than a Hubble time becomes much smaller
than that in the FRW region.

The density contrast across the shock $\delta_{s}\equiv\Delta\rho/\rho_{\rm{FRW}}$,
where $\Delta\rho\equiv\rho_{\rm{shock}}-\rho_{\rm{FRW}}$, is
very large immediately after the wall hits the ambient radiation,
but rapidly drops and becomes ${\cal O}(1)$ in about a Hubble time.
Fig. \ref{fig:The-density-contrast} shows the subsequent evolution
of $\delta_{s}$, which can be approximated as $\delta_{s}(t)\propto t^{-1/2}$.

\begin{figure}
\subfloat[$H_{\rm{b}}=0.05H_{i}$, $H_{\sigma}\approx0.03H_{i},$ and $R_{i}=5H_{i}^{-1}$]{\includegraphics[scale=0.17]{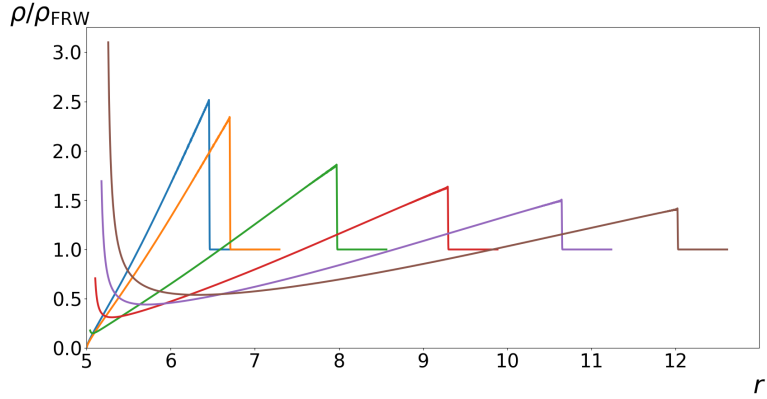}
\captionsetup{justification=justified,singlelinecheck=false}

}

\subfloat[$H_{\rm{b}}=0.75H_{i}$, $H_{\sigma}\approx0.02H_{i},$ and $R_{i}=10H_{i}^{-1}$]{\includegraphics[scale=0.17]{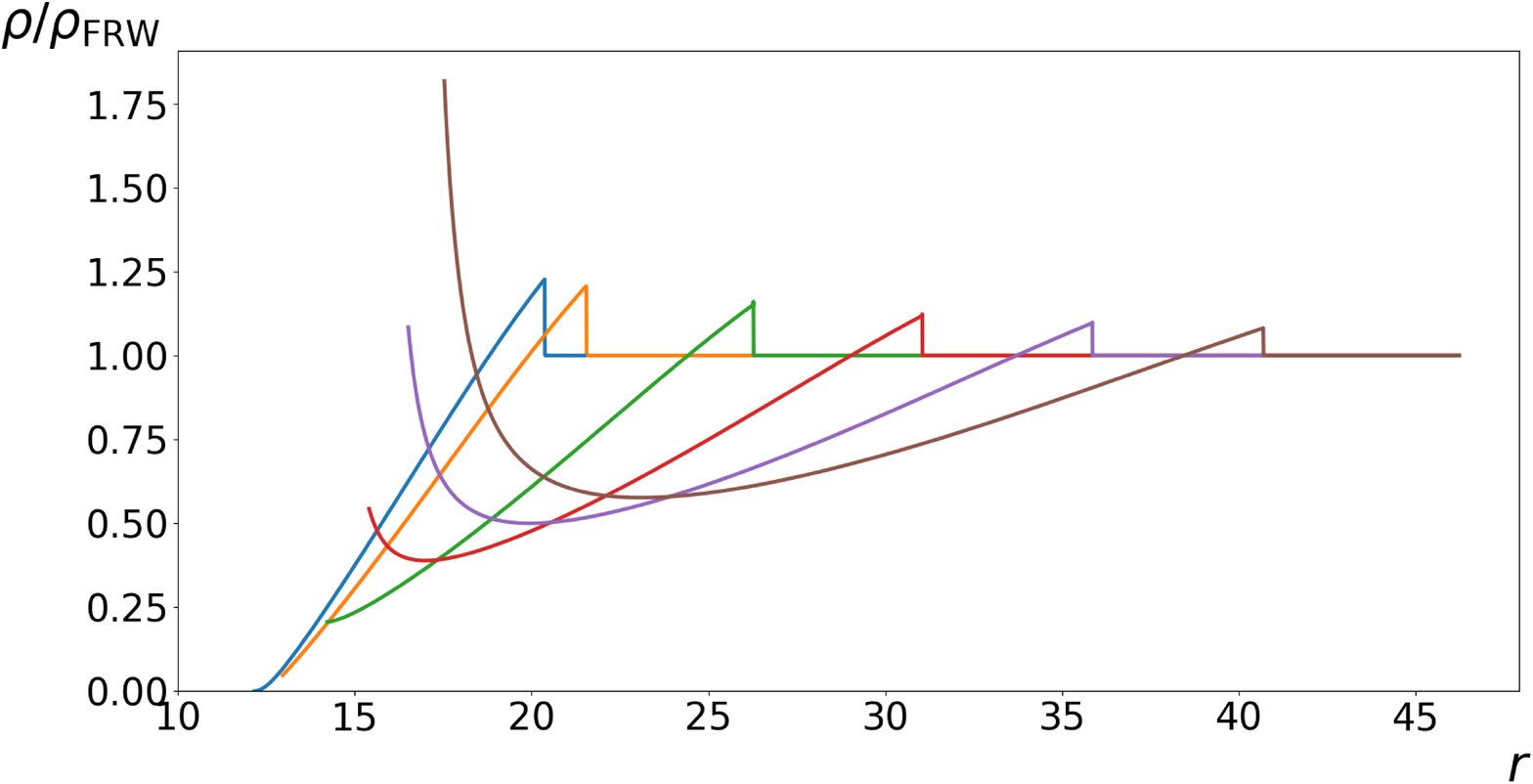}
\captionsetup{justification=justified,singlelinecheck=false}

}

\caption{\label{fig:The-radiation-energy-1}The radiation energy density $\rho$
as a function of the comoving radius $r$ for a subcritical (upper
panel) and a supercritical (lower panel) bubble at different moments
before and after black hole formation. In both plots, the first (blue) and the second (orange) density profiles respectively correspond to moments right before and after the black hole is formed.   After black hole formation, the shock continues to diminish and the density deficit in the
black hole vicinity is gradually filled with radiation. We
cut off the black hole region at the apparent horizon in order to avoid simulation breakdown. In the subcritical case, the apparent horizon arises at the wall; while in the supercritical case, it appears at the wormhole throat. At the first moment (blue) in the second plot, we have already removed the wall and a surrounding layer to avoid simulation breakdown due to the inflating wall.}
\end{figure}

As the shock propagates outwards, the empty layer created by the bubble
impact is gradually filled with radiation. In Fig. \ref{fig:The-radiation-energy-1}
we show the radiation density profile at several moments right before
and after black hole formation, both for a subcritical and a supercritical
bubble. In both cases, in a few Hubble times after the black
hole is formed, it is surrounded by a nearly uniform radiation background.

\subsection{Black hole mass}

\subsubsection{Subcritical bubble}

\begin{figure}
\includegraphics[scale=0.2]{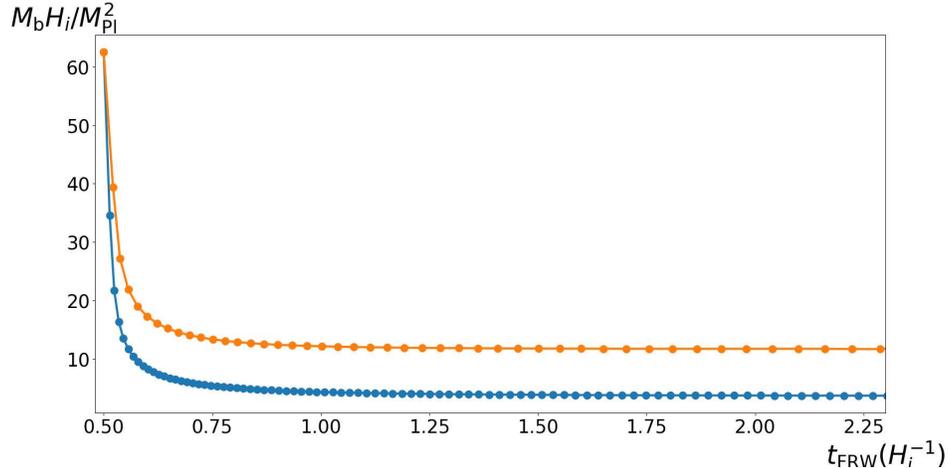}

\caption{\label{fig:Evolution-of-bubble}The early evolution
of the Misner-Sharp mass for a subcritical (blue; $H_{\sigma}\approx0.03H_{i}$,
$H_{\rm{b}}=0.05H_{i}$ and $R_{i}=5H_{i}^{-1}$) and a supercritical
(orange; $H_{\sigma}\approx0.05H_{i}$, $H_{\rm{b}}=0.25H_{i}$
and $R_{i}=5H_{i}^{-1}$) bubble. In both cases, the bubble loses most of
its energy within a time scale much smaller than the Hubble time $H_{i}^{-1}$.
The mass continues to change due to the radiation pressure, but in
about a Hubble time this effect becomes negligible, and the Misner-Sharp mass approaches the conserved mass parameter ${\cal M}_{\rm b}$.  In the subcritical case the initial black hole mass is $M_{\rm bh}\sim {\cal M}_{\rm b}$, while for supercritical bubbles $M_{\rm bh}$ is not simply related to ${\cal M}_{\rm b}$.}
\end{figure}

The Misner-Sharp mass of the bubble decreases dramatically within
a time scale $\Delta t\ll t_{i}$ due to momentum transfer to radiation.
At later times, the mass continues to change due to the radiation
pressure, but it approaches a constant after about one Hubble time
(Fig. \ref{fig:Evolution-of-bubble}). In the subcritical case, the
bubble radius $R_{\rm{w}}$ reaches a maximum and then decreases.
An apparent horizon is formed when $\Theta_{\rm{out}}=0$ at the
wall. We regard this as a signal of black hole formation. The black
hole mass can be estimated as the Misner-Sharp mass at the apparent
horizon.

The black hole masses obtained from the simulations are compared with
the analytic estimate (\ref{GM}) in Table. \ref{tab:BH-masses-for}.
We see that Eq.~(\ref{GM}) gives a good estimate\textbf{ }within
a factor of 2. As expected, in most examples the actual black hole
mass is lower than estimated, because of the radiation pressure during
the expansion phase of the bubble. In some examples, however, the
actual mass is slightly higher. This is because the expansion phase
in these cases was very short.

At later times, the black hole mass grows by accretion of radiation,
but the resulting mass increase is no more than by a factor of 2 \cite{DGV}.

\begin{table}
\begin{tabular}{|c|c|c|c|}
\hline 
Parameters$(H_{i})$  & $R_{i}(H_{i}^{-1})$  & $M_{\rm{est}}(M_{\rm{Pl}}^{2}H_{i}^{-1})$  & $M_{\rm{bh}}(M_{\rm{Pl}}^{2}H_{i}^{-1})$\tabularnewline
\hline 
\hline 
\multirow{4}{*}{$H_{\rm{b}}=0.05$, $H_{\sigma}\approx0.03$} & 2  & 0.5  & 0.3\tabularnewline
\cline{2-4} 
 & 3  & 1.7  & 1.0\tabularnewline
\cline{2-4} 
 & 4  & 3.9  & 2.1\tabularnewline
\cline{2-4} 
 & 5  & 7.6  & 3.7\tabularnewline
\hline 
$H_{\rm{b}}=0.25$, $H_{\sigma}\approx0.03$  & 2  & 0.8 & 0.7\tabularnewline
\hline 
$H_{\rm{b}}=0.5$, $H_{\sigma}\approx0.01$  & 1.8  & 0.8  & 0.9\tabularnewline
\hline 
$H_{\rm{b}}=0.75$, $H_{\sigma}\approx0.02$  & 1  & 0.3 & 0.3\tabularnewline
\hline 
\end{tabular}

\caption{\label{tab:BH-masses-for}Black hole masses for six subcritical bubbles.
$M_{\rm{est}}$ is the estimate given by Eq. \eqref{GM}, and $M_{\rm{bh}}$
is the simulation result. }
\end{table}

\subsubsection{Supercritical bubble}

\begin{figure}
\includegraphics[scale=0.18]{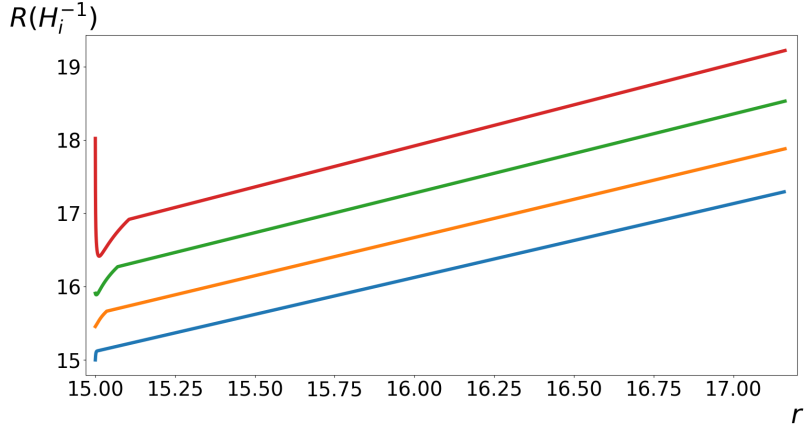}

\caption{\label{fig:Area-radius-}The area radius $R(r)$ at different times
for a supercritical bubble with $H_{\rm{b}}=0.75H_{i}$, $H_{\sigma}\approx0.02H_{i}$
and $R_{i}=15H_{i}^{-1}.$ The bottom blue curve is the initial
profile of $R$. In the unperturbed FRW region, $R\propto r.$ A
local minimum develops with time, indicating the formation of a wormhole
throat. The wall inflates away exponentially afterwards, so $R$ grows
sharply near the wall.}
\end{figure}

In the supercritical case, a wormhole develops outside of the bubble
wall, and the bubble starts to inflate. In Fig. \ref{fig:Area-radius-}
we show the area radius $R$ as a function of the comoving radius
$r$ at several successive moments of time. We see that $R(r)$ develops
a minimum outside of the wall, signaling the formation of a wormhole.
Since the bubble is rapidly expanding, the radius $R(r)$ grows sharply
towards the wall.

In order to simulate the evolution of the region near the bubble wall,
high resolution is needed. However, since the bubble grows supersonically
away from the exterior region and thus gets detached from the fluid,
we cut off the wall as well as a layer immediately outside, so as
to prevent $R(r)$ from changing steeply near the inner boundary and
avoid simulation breakdown. This excision does not affect the evolution
of the exterior region.

The black hole formation is signaled by the horizon bifurcation point,
where $\Theta_{\rm{out}}=\Theta_{\rm{in}}=0$. (This is the point
at the intersection of the two apparent horizon lines in the conformal
diagram of Fig. \ref{sup}.) At this point two black holes of equal
mass are formed, one for the observer in the baby universe and the
other in the exterior FRW universe.

The two black holes start with identical masses, but the masses can
grow later by accretion and do not have to remain equal. The mass
accretion on the exterior black hole has been studied in Ref.~\cite{DGV}
in the domain wall scenario, with the conclusion that it increases
the black hole mass by approximately a factor of 2. For a large supercritical
bubble, the perturbation caused by the shock should mostly
dissipate by the time of black hole formation, and we expect that
the accretion process will be very similar to the domain wall case,
with similar result.

\begin{figure}
\includegraphics[scale=0.23]{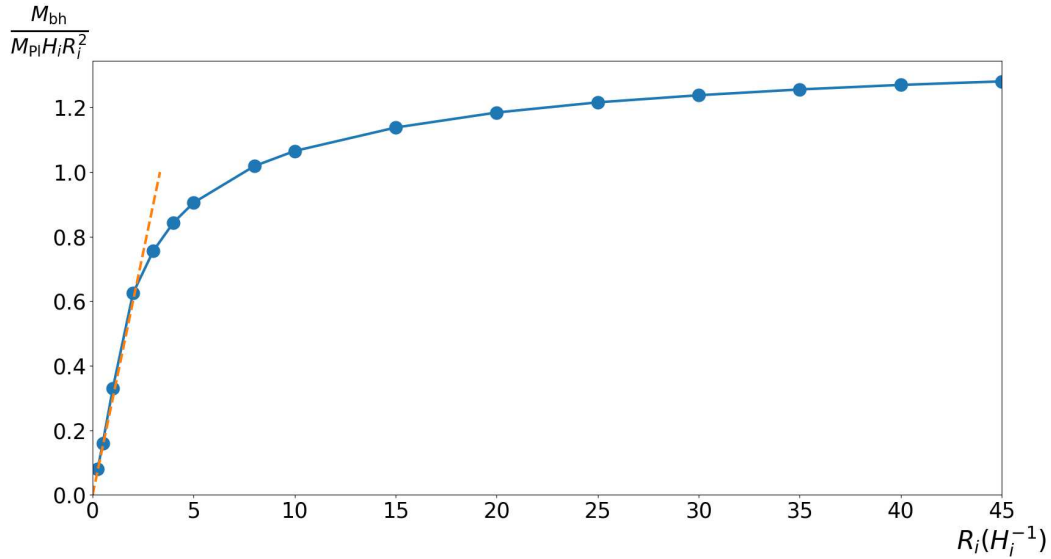}

\caption{\label{fig:Blue-dots-are}Black hole mass as a function of $R_i$. Blue dots are $M_{\rm{bh}}/M_{{\rm Pl}}^{2}H_{i}R_{i}^{2}$
for bubbles with $H_{i}R_{i}=$ 0.25, 0.5, 1, 2, 3, 4, 5, 8, 10, 15, 20, 25, 30, 35, 40 and 45. 
In all cases, $H_{\sigma}\approx0.02H_{i}$ and $H_{\rm{b}}=0.75H_{i}$.
For these parameter values, the critical radius is $R_{\rm{cr}}\sim1.3H_{i}^{-1}$
and the transition radius $R_{*}\sim3H_{i}^{-1}.$  The orange
dashed line shows our estimate for subcritical cases. We can see that
the estimate works well even for the supercritical case with $R_{i}=2H_{i}^{-1}$.}
\end{figure}

In Fig.~\ref{fig:Blue-dots-are} we plotted the ratio $M_{\rm{bh}}/M_{{\rm Pl}}^{2}H_{i}R_{i}^{2}$
for a range of values of $R_{i}$. We see that as $R_{i}$ increases,
the ratio approaches a constant ${\cal O}(1)$, 
\begin{equation}
M_{\rm{{\rm bh}}}\sim M_{{\rm Pl}}^{2}H_{i}R_{i}^{2}.\label{MRi}
\end{equation}
Thus, for large values of $R_{i}$ the bound (\ref{bound}) is nearly
saturated, as in the domain wall scenario.

The data points with $R_{i}<H_{i}^{-1}$ correspond to subcritical regime,
where we expect $M_{\rm{bh}}\sim\kappa M_{\rm{Pl}}^{2}R_{i}^{3}$
with $\kappa$ defined from Eq.(\ref{GM}),
\begin{equation}
\kappa\equiv\frac{1}{2}H_{\rm{b}}^{2}+2H_{\sigma}\left(\sqrt{H_{i}^{2}-H_{\rm{b}}^{2}}-H_{\sigma}\right).
\end{equation}
For the parameter values in Fig.~\ref{fig:Blue-dots-are}, $\kappa\sim0.3H_{i}^{2}$. This estimate is in agreement with the simulation results.

In conclusion, we can roughly approximate our results by setting 
\begin{equation}
M_{{\rm bh}}\sim M_{{\rm Pl}}^{2}\begin{cases}
\kappa R_{i}^{3} & M<M_{*}\\
H_{i}R_{i}^{2} & M>M_{*},
\end{cases}.\label{approxM}
\end{equation}
where the transition mass 
\begin{equation}
M_{*}\sim\frac{M_{{\rm Pl}}^{2}H_{i}^{3}}{\kappa^{2}}
\end{equation}
corresponds to $R_{*}\sim H_{i}/\kappa$.

Parameters used in Fig.~\ref{fig:Blue-dots-are} give $R_{*}\sim3H_{i}^{-1}$.
In this case $M_{*}$ is not much different from $M_{\rm{cr}},$
but for other parameter values these two masses can be rather different.
For example, if the first term in $\kappa$ dominates, we have $M_{*}/M_{{\rm cr}}\sim(H_{i}/H_{\rm b})^{3}$,
which can be large for $H_{\rm{b}}\ll H_{i}$. Restricted by the
capability of our simulation, we did not further explore the transition
regime $M_{{\rm cr}}\lesssim M_{\rm bh} < M_{*}$.

\section{Black hole mass spectrum}

In earlier sections we have shown how a black hole could be formed
by a vacuum bubble after inflation and how its mass is related to
the initial bubble radius $R_{i}$. During inflation the universe
expands by a huge factor, so the bubble radii spread over a large
range of scales. In this section we calculate the distribution of
$R_{i}$ and find the resulting PBH mass spectrum. The calculation
follows closely that in Ref.~\cite{GVZ}.

\subsection{Size distribution of bubbles}

To simplify the notation, in this subsection we use $R$ to denote the bubble radius during inflation.

The background spacetime during inflation can be described by a flat
de Sitter metric, 
\begin{equation}
ds^{2}=-dt^{2}+a^{2}(t)d{\bf{x}}^{2}
\end{equation}
with  $a(t)=H_{i}^{-1}\exp(H_{i}t)$. Let $t_{\rm{n}}$
be the bubble nucleation time. For simplicity, we assume that the
bubble nucleates with a negligible radius compared to $H_{i}^{-1}$;
then the bubble worldsheet is well approximated by the future light cone of the nucleation point, 
\begin{equation}
R(t)\approx H_{i}^{-1}\left[e^{H_{i}(t-t_{\rm{n}})}-1\right].
\end{equation}

The number of bubbles that materialize in a coordinate interval $d^{3}{\bf{x}}$
and time interval $dt_{\rm{n}}$ is 
\begin{equation}
dN=\lambda H_{i}^{4}e^{3H_{i}t_{\rm{n}}}d^{3}{\bf{x}} dt_{\rm{n}},
\end{equation}
where $\lambda$ is the bubble nucleation rate per Hubble spacetime
volume $H_{i}^{-4}$.

The number density of bubbles having radius in the interval $(R,R+dR)$
at time $t$ is 
\begin{equation}
dn(t)\equiv\frac{dN}{dV}=\lambda\frac{dR}{\left(R+H_{i}^{-1}\right)^{4}},\label{dNdV}
\end{equation}
where 
\begin{equation}
dV\equiv e^{3H_{i}t}d^{3}{\bf{x}}
\end{equation}
is the physical volume element at $t$.

The distribution (\ref{dNdV}) applies in the range $R\lesssim H_{i}^{-1}e^{\mathcal{N}}$,
where $\mathcal{N}$ is the number of inflationary e-folding. During
inflation, the form of the distribution does not change with time,
except the upper cutoff increases, reaching its maximum at the end
of inflation.

\subsection{Black hole mass distribution}

To simplify the notation, in this and the next subsection we replace $M_{\rm{bh}}$
by $M$.

By Eq. \eqref{dNdV}, the number density of bubbles within the radius
range $(R_{i},R_{i}+dR_{i})$ at the end of inflation is 
\begin{equation}
dn(t_{i})=\lambda\frac{dR_{i}}{\left(R_{i}+H_{i}^{-1}\right)^{4}}.\label{dndR}
\end{equation}
After $t_{i}$ the bubble population is diluted by cosmic expansion,
i.e. 
\begin{equation}
dn(t)=dn(t_{i})\left[\frac{a(t_{i})}{a(t)}\right]^{3},\label{dn(t)}
\end{equation}
where $a(t)\propto t^{1/2}$ is the scale factor and we assume that
the black holes are formed during the radiation era.

We use the standard definition of the mass function 
\begin{equation}
f(M)=\frac{M^{2}}{\rho_{\rm{CDM}}(t)}\frac{dn(t)}{dM},\label{fdef}
\end{equation}
where $\rho_{\rm{CDM}}(t)$ is the mass density of cold dark matter
(CDM). Here $M^{2}dn/dM$ can be interpreted as the mass density of
black holes in the mass range $\Delta M\sim M$. Since the black hole
density and $\rho_{\rm{CDM}}$ are diluted by the cosmic expansion
in the same way, $f(M)$ remains constant in time. The total fraction
of CDM in PBHs can be expressed as 
\begin{equation}
f_{\rm{PBH}}\equiv\frac{\rho_{\rm{PBH}}(t)}{\rho_{\rm{CDM}}(t)}=\int \frac{dM}{M} f(M),\label{fPBH}
\end{equation}
where $\rho_{\rm{PBH}}(t)$ is the PBH mass density.

During the radiation era $(t<t_{{\rm eq}})$, the dark matter density
is of the order 
\begin{equation}
\rho_{\rm{CDM}}(t)\sim\frac{1}{BGt^{2}}\left(\frac{t}{t_{{\rm eq}}}\right)^{1/2}\sim\frac{M_{{\rm Pl}}^{3}}{Bt^{3/2}{\cal M}_{{\rm eq}}^{1/2}},\label{rhoCDM}
\end{equation}
where $B\sim10$ is a constant and ${\cal M}_{{\rm eq}}\sim t_{{\rm eq}}/G\sim10^{17}~M_{\odot}$
is the dark matter mass within a Hubble radius at $t_{\rm{eq}}$.

To find the mass function for our model, we use the simple ansatz
(\ref{approxM}) for $M(R_{i})$. For bubbles with $R_{i}\gg H_{i}^{-1}$,
we can neglect $H_{i}^{-1}$ in Eq.~(\ref{dndR}). Then, for black
holes with $M>M_{*}$ we have  
\begin{equation}
\frac{dn(t)}{dM}\sim\frac{\lambda M_{{\rm Pl}}^{3}}{M^{5/2}t^{3/2}}.\label{dndM}
\end{equation}
and 
\begin{equation}
f(M)\sim B\lambda\left(\frac{{\cal M}_{{\rm eq}}}{M}\right)^{1/2}.
\end{equation}

For $M<M_{*}$, we use $M\propto R_{i}^{3}$, which gives $dn/dM\propto M^{-2}$
and $f(M)={\rm const}$. Hence the resulting mass function has the
form 
\begin{equation}
f(M)\sim B\lambda{\cal M}_{{\rm eq}}^{1/2}\begin{cases}
M_{*}^{-1/2} & M<M_{*}\\
M^{-1/2} & M>M_{*}.
\end{cases},\label{approxf}
\end{equation}

The distribution (\ref{approxf}) becomes inaccurate for black holes
of mass 
\begin{equation}
M\lesssim\kappa M_{\rm{Pl}}^{2}H_{i}^{-3}\equiv M_{H},\label{MH}
\end{equation}
formed by bubbles with $R_{i}\lesssim H_{i}^{-1}$ that nucleated during
the last e-fold of inflation. For $R_{i}\ll H_{i}^{-1}$, Eq. \eqref{M_b} gives
\begin{equation}
GM\sim\left(\frac{1}{2}H_{\rm{b}}^{2}-2H_{\sigma}^2\right)R_{i}^{3}+2H_{\sigma}R_{i}^{2}.\label{GM2}
\end{equation}
If the first term dominates, $f(M) \propto M^{4/3}$; if the second term dominates, then  $f(M)\propto M^{3/2}$. In either case, the mass function decreases relatively fast at $M<M_{H}$, and thus $M_{H}$ plays
the role of a lower cutoff for the distribution (\ref{approxf}).

Another cutoff mechanism is due to shape fluctuations of the bubbles.\footnote{We are grateful to Jaume Garriga for emphasizing this to us.}  At the time of nucleation, bubbles are not perfectly spherical, because of quantum fluctuations.  The amplitude of these fluctuations and their subsequent evolution have been discussed in Refs.~\cite{GV91,GV92}.  When a subcritical bubble collapses, the shape fluctuations grow and may become large before the bubble shrinks to its Schwarzschild radius.  The bubble will then fragment into smaller pieces, which will in turn disintegrate into relativistic particles, so no black hole will be formed.  We show in Appendix B that the corresponding lower bound on the black hole mass is
\begin{equation}
M_{\rm bh}\gtrsim \rho_{\rm b}\left(\frac{\rho_i M_{\rm Pl}}{\rho_{\rm b} \sigma}\right)^{3/2}\equiv M_F.
\end{equation}
Some black holes with $M_{\rm bh} < M_F$ may still be formed from bubbles with atypically small shape fluctuations. We have not explored the mass distribution in this regime.
 
The shape fluctuations have no effect on the evolution of supercritical bubbles, thus the lower bound cannot be larger than $M_{\rm cr}$. Therefore, the mass function (\ref{approxf}) is effectively cut
off at 
\begin{equation}
M_{{\rm min}}\sim \begin{cases}
\max \{M_H,M_F\}, & M_F<M_{\rm cr}\\
M_{\rm cr}, & M_F>M_{\rm cr}
\end{cases}.
\end{equation}

\begin{figure}\includegraphics[scale=0.077]{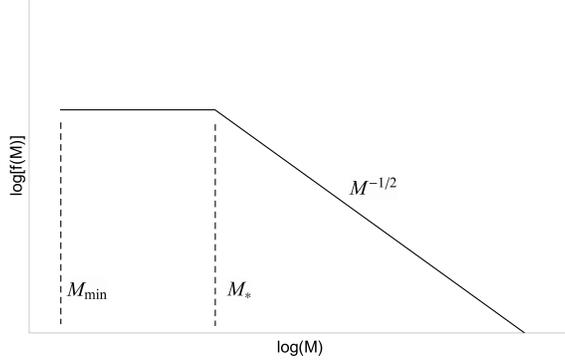}

\caption{\label{fig:A-general-form}A general form of our mass function, Eq.
\eqref{approxf}.  The mass function at $M<M_{\rm min}$ rapidly declines towards zero. Its shape depends on the cutoff mechanism and is not shown here.}
\end{figure}

Depending on the microphysical energy scales $\rho_i$ and $\rho_{\rm b}$, the mass parameters $M_*$ and $M_{\rm min}$ can take a very wide range of values. For example, with $\rho_i$ and $\rho_{\rm b}$ varying between the electroweak and grand unification scale, $M_*$ can be as small as a few grams and can be larger than the mass of the entire observable universe, while $M_{\rm min}$ is restricted to $M_{\rm min} \lesssim 10^{-4} M_\odot$.  On the other hand, if the scale of $\rho_{\rm b}$ is less than the electroweak scale, $M_{\rm min}$ can be much larger.  Here we shall treat $M_*$ and $M_{\rm min}$ as free parameters. A general form of the mass function is illustrated in Fig, \ref{fig:A-general-form}.


By Eq. \eqref{fPBH}, the total mass fraction of dark matter in PBH
is given by 
\begin{equation}
f_{{\rm PBH}}\sim B\lambda\left(\frac{{\cal M}_{{\rm eq}}}{M_{*}}\right)^{1/2}\left[\ln\left(\frac{M_{*}}{M_{{\rm min}}}\right)+1\right].\label{approxfPBH}
\end{equation}

\section{Observational properties and constraints}

Apart from the distinctive mass spectrum, black holes produced by
our mechanism have other interesting properties.

At the time of formation, these black holes are non-rotating. They
may acquire some angular momentum by accretion of matter at later
times, but much of the accretion occurs in the radiation era, within
a few Hubble times after formation, and is likely to be nearly spherically
symmetric. Hence we expect this population of black holes to be very
slowly rotating. It is interesting to note that LIGO observations
suggest low spins for the merging black holes \cite{Hotokezaka:2017dun}
.

Scenarios of PBH formation from large primordial density fluctuations
predict a background of stochastic gravitational waves, imposing significant
constraints on this PBH formation mechanisms. Another stringent constraint
comes from the observational bounds on the $\mu$-distortion of the
CMB spectrum. (For a discussion of these constraints, see, e.g. \cite{Inomata:2016rbd,Garcia-Bellido:2017aan}
and references therein). Our model does not require large initial
fluctuations and is not subject to these constraints.

\begin{figure}
\includegraphics[scale=0.15]{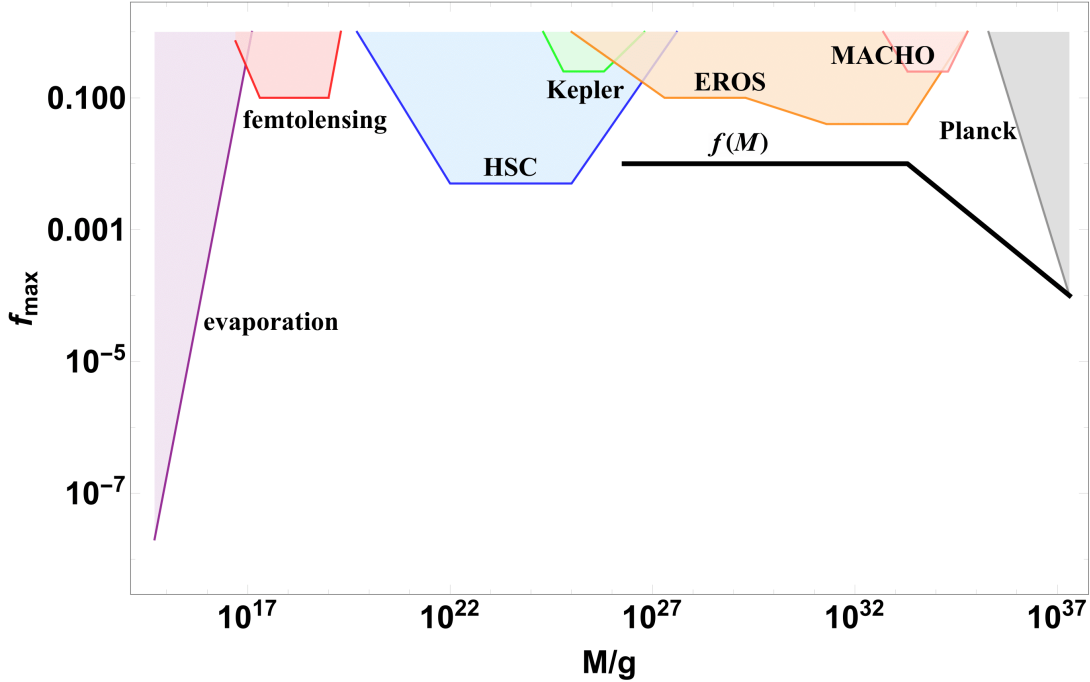}

\caption{\label{PBHcon}A sketch of constraints from different observations on the fraction
of dark matter in PBHs as a function of the PBH mass for a monochromatic
mass distribution. 
More detail can be found in e.g. Ref. \cite{Carr:2016drx,Carr:2017jsz} and references therein.
As an illustration, we also show the PBH distribution for our model with $\lambda\sim 10^{-12}$, $M_* = M_\odot$ and $M_{\rm min} = 10^{-7} ~M_\odot$, which is marginally consistent with the constraints.
 }
\end{figure}

We now turn to observational constraints on our PBH formation model.
Constraints on PBHs in different mass ranges have been extensively
studied in the literature; see, e.g., \cite{Carr:2016drx} for an
up to date review. We have indicated the current constraints in Fig.
\ref{PBHcon}. A very stringent constraint on PBHs with $M\sim M_{{\rm evap}}\sim10^{15}\ \rm{g}$
comes from Hawking evaporation. For our mass function (\ref{approxf}),
a substantial mass fraction in PBHs can be obtained only if the cutoff
mass is $M_{\rm min}>M_{{\rm evap}}$.

\begin{figure}
\includegraphics[scale=0.27]{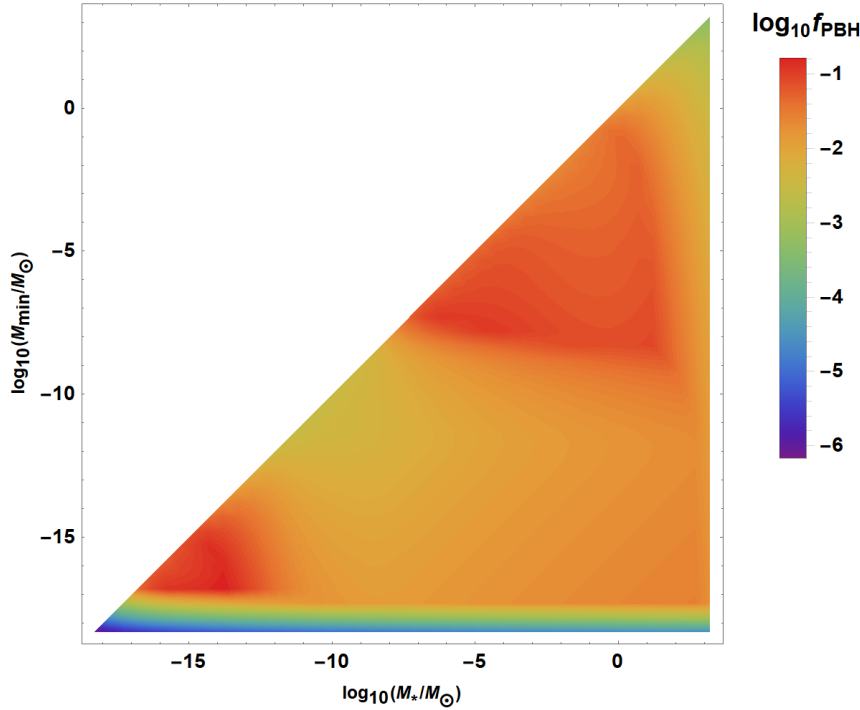}

\caption{\label{PBHplot}Observational upper bound on the fraction of dark
matter in PBHs for different values of the model parameters $M_*$ and $M_{\rm min}$.  It can be seen that in our model PBHs can constitute no more than
$\sim 10\%$ of the dark matter.}
\end{figure}

As discussed in Refs.~\cite{Carr:2016drx,Kuhnel:2017pwq,Carr:2017jsz},
applying the constraints to models like ours, with a broad mass distribution
of PBH, requires a special analysis. For example, observations
like EROS, MACHO and HSC provide bounds $f_{{\rm max}}(M)$ over several
orders of magnitude. For a \textquotedbl{}monochromatic\textquotedbl{}
mass distribution, these bounds simply imply $f(M)<f_{{\rm max}}$,
but for an extended mass distribution they give a somewhat stronger constraint
\cite{Carr:2017jsz} 
\begin{equation}
\int_{M_{1}}^{M_{2}}\frac{dM}{M}\frac{f(M)}{f_{{\rm max}}}<1,\label{extended}
\end{equation}
where $M_{1}<M<M_{2}$ is the range of masses covered by a particular
observation. The resulting upper bound on the fraction of dark matter
in PBHs ($f_{{\rm {PBH}}}$) for different values of the model parameters $M_*$ and $M_{\rm min}$
is shown in Fig. \ref{PBHplot}. It follows from the figure that PBHs in our model can at best constitute of order 10\% of the dark matter.

Recent interest in PBHs is largely inspired by LIGO observations of
gravitational waves emitted by inspiraling black holes with $M\sim10-30~M_{\odot}$.
It has been suggested in \cite{Sasaki:2016jop} that PBHs in this range
of masses with $f(M)\sim 10^{-3}$ could have a sufficient
merger rate to account for LIGO results.\footnote{Ref.~\cite{Bird:2016dcv} suggested that 
a significantly higher PBH density was required, $f(M\sim10~M_{\odot})\sim0.1-1$.  But recent detailed analysis in \cite{Ali-Haimoud:2017rtz} supports the conclusions of \cite{Sasaki:2016jop}.}
This PBH density can in fact be achieved in our scenario. One example with $M_* \sim M_\odot$ and $\lambda \sim 10^{-12}$ is shown in Fig. \ref{PBHcon}.

Another interesting possibility is that PBHs could serve as seeds
for supermassive black holes (SMBH) observed at the galactic centers.
The mass of such primordial seeds should be $M\gtrsim10^{3}~M_{\odot}$
\cite{Duechting:2004dk}, and could be significantly higher.
Their number density at present should be comparable to the density
of large galaxies, $n_{G}\sim0.1~\rm{Mpc}^{-3}$.  The relevant question for our scenario is then: What is the largest PBH we can expect to find in a galaxy?

The number density of PBHs of mass $\sim M$ is approximately given by
\begin{equation}
n(M)\sim\rho_{{\rm CDM}}f(M)/M.
\end{equation}
For $M > M_*$ the mass function depends only on the nucleation rate $\lambda$, and requiring that $n(M) \sim n_G$ we find $M \sim 10^{14}\lambda^{2/3}M_\odot$. For $\lambda \sim 10^{-12}$ (which is the upper bound of $\lambda$), this is $\sim 10^6~M_\odot$, which would certainly be sufficient to seed SMBH.

We thus see that for some values of the parameters black holes produced
by high-energy bubbles can have interesting astrophysical implications.
In particular, they could seed SMBH and could account for LIGO observations.

\section{Conclusions and discussion}

In this paper we used numerical simulations to study primordial
black holes formed by vacuum bubbles created during inflation. 
At the end of inflation the bubbles have a scale-invariant size distribution, and we found in Eq. \eqref{approxM} how the black hole mass is related to the bubble radius.  Bubbles smaller than certain critical size collapse to a Schwarzschild singularity; their mass has been estimated analytically in Ref.~\cite{GVZ}.  Supercritical bubbles, on the other hand, inflate in a baby universe, and \cite{GVZ} only found an upper bound on the black hole mass in this case.  Here we confirmed the estimate of \cite{GVZ} for subcritical bubbles and showed that supercritical bubbles nearly saturate the upper bound on the black hole mass. 

Black holes in this model have a very wide mass distribution, Eq.\eqref{approxf}, stretching over many orders of magnitude.  This distribution has a characteristic mass $M_*$ and has a distinctly different form at $M_{\rm bh} > M_*$ and $M_{\rm bh} < M_*$.  There is also a cutoff mass $M_{\rm min}$ below which $f(M_{\rm bh})$ rapidly declines towards zero.  Depending on the microphysics parameters, the characteristic mass $M_*$ can take a wide range of values, from less than a kilogram to much greater than $M_\odot$.

The distribution at $M_{\rm bh} > M_*$ has the same form, $f(M_{\rm bh})\propto M_{\rm bh}^{-1/2}$, as that predicted in models where black holes are formed by scale-invariant density fluctuations in a radiation-dominated universe \cite{Carr:1975qj}.  Black holes in this scenario are formed on the horizon scale, as in the supercritical regime of our model, so the coincidence of the mass functions is not surprising.  We note, however, that the overdensity required for a horizon-size region of radiation to collapse is $\delta\rho /\rho \sim 1$, and in order to form a substantial number of black holes the rms density fluctuations should be $(\delta\rho/\rho)_{\rm rms} \gtrsim 0.1$, much larger than that indicated by CMB and large-scale structure observations. For this reason, models predicting appreciable black hole formation from density fluctuations assume that the primordial fluctuation spectrum has an enhanced amplitude at relatively small scales (e.g., \cite{Inomata:2016rbd,Garcia-Bellido:2017aan}).

A mass distribution $\propto M_{\rm bh}^{-1/2}$ is also predicted in models closely
related to ours, where PBH are formed by spherical domain walls \cite{GVZ,DGV} or circular loops of cosmic string \cite{Garriga:1992nm} formed during inflation.  The domain wall model also has a characteristic mass $M_*$ above which black holes contain inflating baby universes.  In the string model, the mass spectrum extends to very small masses and the black hole density is severely constrained by the Hawking radiation bound.

Considering the constraints from different observations, we found an upper bound on the fraction
of dark matter in PBHs in our model, shown in Fig. \ref{PBHplot}. PBHs here can constitute no more than about 10\% of the dark matter. Furthermore, we found that the fraction of dark matter
in PBHs with mass $M_{\rm{bh}}\sim10 - 30 ~M_{\odot}$ is $\lesssim 10^{-2},$ which implies
that the black holes detected in LIGO events could be the PBHs of
our model.  With a large number of future merger detections, we should be able to infer the black hole mass spectrum (e.g., \cite{Zevin:2017evb}) and to confirm or rule out our PBH formation mechanism.

We also determined an upper bound on the mass of the largest PBH that one can expect to find in a large galaxy, $M_{\rm{bh}}\lesssim 10^7 ~M_{\odot}$.  This shows that PBHs of our model can serve as seeds of supermassive black holes (SMBH) observed at the galactic centers.  There is in fact a range of parameters for which the model can account for both SMBH and LIGO observations.

Black holes formed by vacuum bubbles can have significant observational effects only if the bubble nucleation rate is relatively high, $\lambda\sim 10^{-12} - 10^{-15}$.  This corresponds to the tunneling action $S\sim 30$, while the typical value is $S\gtrsim 100$.  We note, however, that in the string landscape scenario the false vacuum has a large number $(\sim 100)$ of decay channels, and it seems likely that some of them may have small barriers with a relatively high tunneling probability.

Our analysis in this paper is based on several simplifying assumptions.  To begin with, we assumed that the vacuum energy is instantaneously thermalized at the end of inflation.  In a more realistic model, thermalization may extend over several Hubble times and may be preceded by a period of the inflaton field oscillations characterized by the dust equation of state.  We also assumed that 
the bubble radius at nucleation is much smaller than the Hubble horizon.

These assumptions, however, affect only the low-mass end of the black hole distribution, $M_{\rm{bh}}\lesssim M_{\rm min}$.  Perhaps more consequential is the assumption that thermalized matter particles cannot penetrate the bubble and are reflected from the bubble wall.  This is likely to be true for particles with energies lower than the symmetry breaking scale of the bubble, $\epsilon < \eta_{\rm{b}}, \eta_\sigma$, but at higher energies the wall may be transparent to the particles \cite{Everett:1974bs}.

We also assumed that the bubble wall tension $\sigma$ and the bubble nucleation rate $\lambda$ are constant parameters.  However, this is not generally the case.  When the bubble is formed, the wall tension is determined by the shape of the barrier between the inflating false vacuum of energy density $\rho_i$ and the vacuum in the bubble interior of energy density $\rho_{\rm b}$.  On the other hand, at the end of inflation the barrier is between our vacuum of very low energy density and the bubble interior. This means that the shape of the barrier and the wall tension can change significantly in the course of inflation.

The bubble nucleation rate $\lambda$ may also change during inflation.  As the inflaton field rolls downhill, it moves relative to the minimum at $\rho_{\rm b}$, so the tunneling path (and therefore the tunneling action) are also changing.  As a result the mass distribution of black holes could be significantly modified and could develop a peak at the value of $M_{\rm bh}$ that corresponds to the smallest tunneling action.  We leave the analysis of these possibilities for future research.

\begin{acknowledgments}
This work was supported by the National Science Foundation under grant 1518742.  H.D. was also supported by Burlingame Fellowship at Tufts University.  We are grateful to Jaume Garriga and Andrei Gruzinov for very useful discussions and to Vladyslav Syrotenko for his participation at the early stages of this project. We would also like to thank Bernard Carr for useful comments on the manuscript, and Xiaozhe Hu for helpful advice on some simulation issues. 
\end{acknowledgments}

\appendix
\section{}

In this appendix we use Israel's junction conditions to find the conditions
needed for simulations. We closely follow the method and notation
in \cite{Tanahashi}.

The interior of the vacuum bubble can be described by de Sitter spacetime
with metric

\begin{equation}
ds^{2}=-dt_{\rm{d}}^{2}+a^{2}(dr_{\rm{d}}^{2}+r_{\rm{d}}^{2}d\Omega_{2}^{2}),
\end{equation}
where $a(t_{\rm{d}})=H_{\rm{b}}^{-1}\exp(H_{\rm{b}}t_{\rm{d}})$
with $H_{\rm{b}}$ the Hubble constant.
We assume that outside the bubble
\begin{equation}
ds^{2}=-A^{2}dt^{2}+B^{2}dr^{2}+R^{2}d\Omega_{2}^{2},
\end{equation}
where $A$, $B$ and $R$ are functions of $t$ and $r$.

Let the trajectory of the wall be $(t_{\rm{d}}(\tau),r_{\rm{d}}(\tau))$
or $(t(\tau),r(\tau))$, where $\tau$ is the wall proper time. If
the wall is comoving for an exterior observer, $\partial_{\tau}r=0$,
where $\partial_{\tau}\equiv d/d\tau$. The tangent vector to the
wall hypersurface is $v^{\mu}=(\partial_{\tau}t,\partial_{\tau}r)=(\partial_{\tau}t,0)$,
and $v_{\mu}=(-A^{2}\partial_{\tau}t,B^{2}\partial_{\tau}r)=(-A^{2}\partial_{\tau}t,0)$.
Assuming $\partial_{\tau}t$ to be positive, $v^{\mu}v_{\mu}=-1$
gives $\partial_{\tau}t=A^{-1}.$ Let $\xi^{\mu}$ be a unit vector normal to the wall hypersurface;
then $\xi^{\mu}v_{\mu}=0$ and $\xi^{\mu}\xi_{\mu}=1$, which give
$\xi^{\mu}=(A^{-1}B\partial_{\tau}r,AB^{-1}\partial_{\tau}t)=(0,B^{-1})$
and $\xi_{\mu}=(0,B)$.

We define the brackets $\left[ Q \right] \equiv Q_{\rm{out}}-Q_{\rm{in}}$,
and $\{Q\}\equiv Q_{\rm{out}}+Q_{\rm{in}}$. Here "in" and "out" respectively denote the value of $Q$ right inside and outside the wall. Then Israel's  first junction condition is $[h_{\mu\nu}]=0$, where
$h_{\mu\nu}$ is the induced metric at the wall. The second junction condition is $[K_{\mu\nu}]=8\pi(-S_{\mu\nu}+Sh_{\mu\nu}/2)$,
where $K_{\mu\nu}$ is the extrinsic curvature at the wall, $S_{\mu\nu}=-\sigma h_{\mu\nu}$
is the energy-momentum tensor of the wall, with $\sigma$ the surface
energy density (or tension).

The equation of motion for the wall is given by 
\begin{equation}
\left\{ \xi_{\mu}\frac{Dv^{\mu}}{d\tau}+2\xi^{\mu}\partial_{\mu}\ln R\right\} =-\frac{2}{\sigma}\left[(\rho+p)(u^{\mu}\xi_{\mu})^{2}+p\right],\label{EOM}
\end{equation}
where $u^{\mu}$ is the 4-velocity of the fluid. Inside the bubble
$\rho+p=0$. 

We shall now use the junction conditions and Eq.(A3) to derive some relations that will be useful for setting up the boundary condition for our simulations. 

By the first junction condition, $ar_{\rm{d}}=R$ at the wall. Taking the derivative of $R$ with respect to $\tau$ gives

\begin{equation}
\partial_{\tau}R=\dot{R}\partial_{\tau}t+R^{\prime}\partial_{\tau}r=\frac{\dot{R}}{A}\equiv U.
\end{equation}
Let $V\equiv a\partial_{\tau}r_{\rm{d}}.$ Then $U$ at the wall can be written as

\begin{equation}
U=\partial_{\tau}(ar_{\rm{d}})=r_{\rm{d}}\partial_{\tau}a+a\partial_{\tau}r_{\rm{d}}=H_{\rm{b}}R\sqrt{1+V^{2}}+V.\label{R1}
\end{equation}

The $(\theta,\theta)$ component of the second junction condition
gives 
\begin{equation}
\left[\xi^{\mu}\partial_{\mu}R\right]=-4\pi\sigma R.\label{matching}
\end{equation}
Right outside the wall, we have

\begin{equation}
\left.\xi^{\mu}\partial_{\mu}R\right|_{\rm{out}}=\frac{R^{\prime}}{B}\equiv\Gamma,\label{3}
\end{equation}
while

\begin{align}
\left.\xi^{\mu}\partial_{\mu}R\right|_{\rm{in}} & =H_{\rm{b}}RV+\sqrt{1+V^{2}}.\label{4}
\end{align}
Then by Eq. \eqref{matching}, $\Gamma$ at the wall can be written as

\begin{equation}
\Gamma=H_{\rm{b}}RV+\sqrt{1+V^{2}}-4\pi\sigma R.\label{R2}
\end{equation}

The $(\tau,\tau)$ component of the second junction condition gives
\begin{equation}
\left[ \xi_{\mu}\frac{Dv^{\mu}}{d\tau}\right]=-4\pi\sigma,\label{tautau}
\end{equation}
where $Dv^{\mu}/d\tau=\partial_{\tau}v^{\mu}+\Gamma_{\lambda\sigma}^{\mu}v^{\lambda}v^{\sigma},$
with $\Gamma_{\lambda\sigma}^{\mu}$ the Chistoffel symbols for the 4-spacetime. Right outside the wall,

\begin{equation}
\left.\xi_{\mu}\frac{Dv^{\mu}}{d\tau}\right|_{\rm{out}}=B\Gamma_{00}^{1}v^{0}v^{0}=\frac{A^{\prime}}{AB}.\label{tautauout}
\end{equation}
By Eqs. \eqref{tautau} and \eqref{tautauout},

\begin{equation}
\frac{A^{\prime}}{AB}-\left.\xi_{\mu}\frac{Dv^{\mu}}{d\tau}\right|_{\rm{in}}=-4\pi\sigma.\label{A1}
\end{equation}
On the other hand, by Eqs. \eqref{EOM}, \eqref{matching}, \eqref{3}, \eqref{4} and \eqref{tautauout}, we have

\begin{equation}
\frac{A^{\prime}}{AB}+\left.\xi_{\mu}\frac{Dv^{\mu}}{d\tau}\right|_{\rm{in}}=-\frac{2}{\sigma}\left[ p\right]-\frac{4\Gamma}{R}-8\pi\sigma.\label{A2}
\end{equation}
Combining Eqs. \eqref{A1} and \eqref{A2} gives

\begin{equation}
A^{\prime}=-AB\left(\frac{\left[ p\right]}{\sigma}+\frac{2\Gamma}{R}+6\pi\sigma\right).\label{Aprime-1}
\end{equation}
This is used as the boundary condition in the simulations.

We can also obtain an explicit form of the equation of motion for the wall. It can be shown that

\begin{equation}
\left.\xi_{\mu}\frac{Dv^{\mu}}{d\tau}\right|_{\rm{in}}=\frac{\partial_{\tau}V}{\sqrt{1+V^{2}}}+H_{\rm{b}}V.\label{EOM2}
\end{equation}
Then by Eqs. \eqref{R2}, \eqref{A1} , \eqref{A2} and \eqref{EOM2},  the equation of motion of the
wall is

\begin{equation}
\frac{\partial_{\tau}V}{\sqrt{1+V^{2}}}=-3H_{\rm{b}}V-\frac{2}{R}\sqrt{1+V^{2}}+6\pi\sigma-\frac{\left[ p\right]}{\sigma}.
\end{equation}

\section{}
Vacuum bubbles can deviate from spherical shape due to quantum fluctuations.  The unperturbed worldsheet of the bubble wall is a $(2+1)$-dimensional de Sitter space with a Hubble parameter
\begin{equation}
{\tilde H}=\epsilon/3\sigma,
\end{equation}
where $\epsilon$ is the difference of vacuum energy densities outside and inside the bubble, $\epsilon =\rho_i -\rho_{\rm b}$
and $\sigma$ is the tension of the bubble wall.  The magnitude of fluctuations of the bubble radius was estimated in Ref.~\cite{GV92}:
\begin{equation}
\delta R \approx \left(\frac{\tilde H}{3\pi^2 \sigma}\right)^{1/2} =\frac{\sqrt{\epsilon}}{3\pi\sigma}.
\end{equation}
As the bubble expands, $\delta R$ remains constant, so the ratio $\delta R/R$ decreases and the bubble becomes more and more spherical. Here $R$ is the bubble radius.

This analysis, however, did not account for gravitational effects.  When the bubble radius gets larger than the de Sitter horizon of the exterior inflating universe, $R>H_i^{-1}$, we expect that the shape of the bubble "freezes" and it is conformally stretched afterwards.\footnote{More exactly, we expect fluctuations of wavelength $\lambda$ to freeze when $\lambda$ gets larger than $1/H_i$.  But here we are interested in the lowest multipoles, so $\lambda$ is comparable to the bubble radius.}
The shape fluctuations are then given by
\begin{equation}
\frac{\delta R}{R} \sim \frac{\sqrt{\epsilon}H_i}{3\pi\sigma}.
\end{equation}
Assuming that $\epsilon\sim\rho_i$, this is of the order
\begin{equation}
\frac{\delta R}{R} \sim \frac{\rho_i}{\sigma M_{\rm Pl}}.
\end{equation}

Widrow \cite{Widrow} studied perturbations on collapsing domain walls and found that the fluctuation $\delta R$ remains approximately constant in the course of collapse.  We will show that this also holds for collapsing bubbles later in this Appendix.

A black hole is formed if $\delta R < 2GM_{\rm bh}$.  Let us first assume that the bubble energy at the moment of maximal expansion $(R=R_{\rm max})$ is dominated by the interior vacuum energy,
\begin{equation}
M_{\rm bh} \sim \rho_{\rm b} R_{\rm max}^3.
\label{Mbh}
\end{equation}
Requiring that $\delta R \lesssim GM_{\rm bh}$, we have
\begin{equation}
\delta R \sim \frac{\rho_i}{\sigma M_{\rm Pl}} R_{\rm max} \lesssim G\rho_{\rm b} R_{\rm max}^3
\end{equation}
and
\begin{equation}
R_{\rm max}\gtrsim \left(\frac{\rho_i M_{\rm Pl}}{\rho_{\rm b} \sigma}\right)^{1/2}.
\end{equation}
Substituting this into (\ref{Mbh}) we obtain
\begin{equation}
M_{\rm bh}\gtrsim \rho_{\rm b}\left(\frac{\rho_i M_{\rm Pl}}{\rho_{\rm b} \sigma}\right)^{3/2}\equiv M_F.
\end{equation}
If $\rho_i$, $\rho_{\rm b}$ and $\sigma$ are characterized by the same energy scale $\eta$, this gives
\begin{equation}
M_{\rm bh}\gtrsim \frac{M_{\rm Pl}^{3/2}}{\eta^{1/2}}.
\end{equation}
This is a rather weak constraint.  For example, if $\eta$ is the electroweak scale, it gives $M_{\rm bh} > 1$~kg.
On the other hand, the cutoff mass $M_F$ can be relatively large when $\rho_{\rm b} \ll \rho_i$.  

\subsection*{Bubble collapse}

Let us now show that the evolution of perturbations on a collapsing bubble is similar to that on a collapsing domain wall.

Let us first consider collapse of a spherical bubble.  We start with the conserved mass,
\begin{equation}
{\cal M}_{\rm{b}}=\frac{4\pi}{3}\rho_{\rm b} R^{3}+4\pi{\sigma}R^{2}\sqrt{1+\dot{R}^{2}-H_{\rm b}^{2}R^{2}}- 8\pi^2 G{\sigma}^{2}R^{3},
\label{M_bb}
\end{equation}
where the overdot stands for a derivative with respect to proper time $\tau$.
Suppose the bubble is initially at rest, ${\dot R}=0$, at the maximal expansion radius, $R=R_{\rm max}$.  We will be interested in the asymptotic behavior, when $R \ll R_{\rm max}$.  

In this regime, the terms proportional to $R^3$ on the right hand side of (\ref{M_bb}) become negligible and $|{\dot R}|\gg 1$, so Eq.~(\ref{M_bb}) reduces to
\begin{equation}
{\cal M}_{\rm{b}} \approx -4\pi{\sigma}R^{2} \dot{R} ,
\label{M2}
\end{equation}
where the "-" sign is chosen because we are considering the collapse.  

The solution of Eq.~(\ref{M2}) is
\begin{equation}
R(\tau)=\left(-\frac{3{\cal M}_{\rm{b}}\tau}{4\pi\sigma}\right)^{1/3}.
\label{Rtau}
\end{equation}
Here, we choose the origin of $\tau$ so that $\tau=0$ at the moment of collapse; then $\tau <0$ prior to the collapse.  It follows from (\ref{Rtau}) that the total proper time it takes for the bubble to collapse is
\begin{equation}
\tau\sim \frac{\sigma R_{\rm max}^3}{\cal M_{\rm{b}}}\lesssim \frac{\sigma}{\rho_{\rm b}}.
\label{tau}
\end{equation}

Fluctuations on the collapsing bubble are described by a scalar field $\phi$ living at the bubble wall and satisfying the equation
\begin{equation}
-\nabla^2 \phi +\left({\cal R}-\frac{\rho_{\rm b}^2}{\sigma^2}\right) \phi=0 .
\label{phieq}
\end{equation}
The field has a tachyonic mass, $m_\phi^2 = - {\rho_{\rm b}^2}/{\sigma^2}$, and a non-minimal coupling to the 3-curvature ${\cal R}$ on the wall worldsheet,
\begin{equation}
{\cal R}=\frac{2}{R^2}\left(1+{\dot R}^2+2R{\ddot R}\right).
\end{equation}
The worldsheet metric can be written as
\begin{equation}
ds^2 = -d\tau^2 +R^2(\tau) d\Omega^2.
\end{equation}
Then, with the solution (\ref{Rtau}) the 3-curvature becomes
\begin{equation}
{\cal R}=\frac{2}{R^2}-\frac{2}{3\tau^2} \approx -\frac{2}{3\tau^2},
\label{calR}
\end{equation}
where the last approximation applies for $R\ll R_{\rm max}$.

The tachyonic mass could in principle lead to an instability.  But the timescale for such an instability to develop is $\Delta\tau > |m_\phi|^{-1}$, and we see from (\ref{tau}) that there is not enough time.  In fact, the mass term in Eq.~(\ref{phieq}) is negligible to compared to the curvature term for $R\ll R_{\rm max}$.
Neglecting this term and using the approximation (\ref{calR}), we rewrite Eq.~(\ref{phieq}) as
\begin{equation}
{\ddot\phi}+\frac{2}{3\tau} {\dot\phi} -\frac{2}{3\tau^2}\phi =0.
\end{equation}
This has solutions $\phi\propto\tau^\alpha$ with $\alpha=1,-2/3$.  The dominant solution is
\begin{equation}
\phi\propto \tau^{-2/3}\propto R^{-2}.
\label{phiR}
\end{equation}

It was shown in \cite{GV91} that the rms fluctuation of the bubble wall $\delta R$ is related to $\phi$ as
$\delta R \sim \phi /\gamma$, where $\gamma \propto R^{-2}$ is the Lorentz factor of the wall.  (The factor of $\gamma$ accounts for the effect of Lorentz contraction of bubble fluctuations.)  Then Eq.~(\ref{phiR}) tells us that the fluctuation amplitude does not change in the course of collapse:
\begin{equation}
\delta R\approx {\rm const}.
\end{equation}
This agrees with the result obtained by Widrow \cite{Widrow} for collapsing domain walls.  The agreement between the two cases is not surprising, since the vacuum energy density $\rho_{\rm b}$ becomes dynamically unimportant at $R\ll R_{\rm max}$.

\end{document}